\documentclass[
pre, 
twocolumn, superscriptaddress, a4paper, showpacs]{revtex4}

\usepackage{amsfonts,amsmath,amssymb}
\usepackage{graphicx}

\DeclareMathOperator{\bFP}{\widehat{\mathcal{L}}_{FP_B}}
\DeclareMathOperator{\fFP}{\widehat{\mathcal{L}}_{FP_F}}
\DeclareMathOperator{\bfp}{\widehat{\ell}_{FP_B}}
\DeclareMathOperator{\ffp}{\widehat{\ell}_{FP_F}}
\DeclareMathOperator{\J}{J}%
\DeclareMathOperator{\JJ}{\widehat{\mathbf{J}}}
\newtheorem{proposition}{Proposition}

\begin{document}

\title{Boundary singularities and boundary conditions for the Fokker-Planck equations}
\author{Ihor Lubashevsky}
    \email{ialub@fpl.gpi.ru}
    \affiliation{Theory Department, A.M. Prokhorov General Physics Institute, Russian
    Academy of Sciences, \\ Vavilov Str. 38, 119991 Moscow, Russia}
\author{Rudolf Friedrich}
    \email{fiddir@uni-muenster.de}
    \affiliation{Institut f\"ur Theoretische Physik,
    Westf\"alische Wilhelms-Universit\"at M\"unster, D-48149 M\"unster,
    Germany}
\author{Reinhard Mahnke}
    \email{reinhard.mahnke@uni-rostock.de}
   \affiliation{Institut f\"ur Physik, Universit\"at Rostock, D-18051 Rostock,
   Germany}
\author{Andrey Ushakov}
    \email{ushakov-an1986@rambler.ru}
    \affiliation{Moscow Technical University of Radioengineering, Electronics, and Automation,
        \\ Vernadsky pros., 78, 117454 Moscow, Russia}
\author{Nikolay Kubrakov}
    \affiliation{Theory Department, A.M. Prokhorov General Physics Institute, Russian
    Academy of Sciences, \\ Vavilov Str. 38, 119991 Moscow, Russia}
\date{\today}
%
\begin{abstract}
The boundary conditions for the Fokker-Planck equations, forward and
backward ones are directly derived from the Chapman-Kolmogorov equation for
$M$-dimensional region with boundaries. The boundaries are assumed, in
addition, to be able to absorb wandering particles or to give rise to fast
surface transport. It is demonstrated that the boundaries break down the
symmetry of random walks in their vicinity, leading to the boundary
singularities in the corresponding kinetic coefficients. Eliminating these
singularities we get the desired boundary conditions. As it must be the
boundary condition for the forward Fokker-Planck equation matches the mass
conservation.
\end{abstract}

\pacs{02.50.Cw, 02.50.Ga, 02.60.Lj, 05.60.Cd}

\maketitle
\section{The Chapman-Kolmogorov and Fokker-Planck equations. Effect of the medium boundaries}

As is well-known \cite{Gardiner,Risken} Markovian stochastic processes are
completely determined by their transition probabilities which obey the
Chapman-Kolmogorov equation. The Kra\-mers-Moyal expansion can be used to
determine the Fok\-ker-Planck equation by specifying drift vector and diffusion
tensor based on the assumption of vanishing higher order Kramers-Moyal
coefficients.

Usually, the Fokker-Planck equations are derived implicitly assuming that the
phase space of the stochastic variables under consideration extends to infinity
so that so-called natural boundary conditions apply. If stochastic processes in
a finite region of phase space are considered, boundary conditions are
introduced a posteriori based on apparent physical arguments leading to the
notion of a reflecting barrier, characterized by a vanishing normal component
of the probability current, an absorbing barrier, where the probability
distribution has to vanish, and boundary conditions at a discontinuity, where
probability distributions and the normal components of the probability current
have to be continuous. No attempts, so far, have been made to derive the
Fokker-Planck equation simultaneously with appropriate boundary conditions from
the Chapman-Kolmogorov equation.

It is quite evident that boundaries can strongly influence the stochastic
motion of a particle in various ways depending on the microscopic interactions.
As an example we mention a boundary formed by a fast diffusion layer. In such a
thin layer particles are able to diffuse in the directions tangential to the
boundary on a fast time scale, whereas in the bulk the particles behavior
should accurately be described by the Fokker-Planck equation. The theoretical
treatment of the particle diffusion requires a formulation of consistent
boundary conditions which match the internal Fokker-Planck behavior to the
stochastic properties of the boundary layer.

So it could be desirable to have a technique of \textit{deriving} the boundary
conditions applying directly to the manner of the region boundaries affecting
stochastic processes. In this respect we note paper~\cite{BBF} devoted to the
general description of random processes near boundaries causing deterministic
jumps, paper~\cite{AH} deriving boundary conditions for the Fokker-Planck
equation describing coupled transport of photons and electrons, a serious of
papers \cite{SL,RKN,MS} dealing with boundary conditions for the
advection-diffusion problem combining the Boltzmann and Fokker-Planck equations
and their numerical implementation, and also work \cite{S} developing diffusion
models for molecular transport across membranes via ion channels and wider
pores in terms of random walks affected by boundaries with complex properties.
In addition paper~\cite{H} actually constructs the absorbing boundary as a
limit transition of an infinite space with half-spaces different in properties
substantially and work~\cite{WW} implements boundary conditions for Wiener
processes in path integrals. Papers~\cite{G,HG} develop a rather sophisticated
moment technique for tackling the Fokker-Planck equation with mixed boundary
conditions based on a special moment truncation scheme.

In the present paper we shall extend the method of deriving the Fokker-Planck
equation from the Chapman-Kolmogorov equation in such a way that simultaneously
consistent boundary conditions can be formulated. Our approach is based on
introducing physical models for the stochastic behavior close to the boundary.
We explicitly demonstrate that boundaries break the symmetry of the random
forces leading to boundary singularities in the Kramers-Moyal expansion. The
cancelation of these singularities yields the appropriate boundary conditions.
We explicitly derive the boundary conditions for a reflecting or absorbing
barrier as well as boundaries with mixed properties, and describe the general
procedure for the derivation of the boundary conditions for the case of the
fast diffusion layer. It should be noted that a similar anomalous effect of the
region boundaries on random processes was analyzed in papers~\cite{LS,O,PB} in
numerical implementation of Wiener processes in their vicinity. Besides,
paper~\cite{M} applies also to the concept of the symmetry breakdown caused,
however, by external fields in constructing a generalized master equation for
the classic and anomalous diffusion processes.

In principle the present approach can be extended to anomalous transport
phenomena, e.g., sub- and super-diffusion, which are modeled by fractional
diffusion operators. It is well-known that the formulation of boundary
conditions for these processes is still a challenging problem although several
approaches have been developed \cite{BBH,LNH,S1,KPN}. The procedure outlined in
the present paper might be helpful in formulating appropriate boundary
conditions for these more involved processes.

The paper is organized as follows. Section~\ref{sec0} presents the problem
under consideration and sketches out deriving two types of the Fokker-Planck
equations based on the general Chapman-Kolmogorov equation for Markovian
processes. Finally it formulates the problem of the corresponding boundary
conditions and derives the general expressions that should be fulfilled at the
boundaries of medium. Section~\ref{sec1} discuses the types of medium
boundaries and their properties to be taken into account. Section~\ref{sec2}
introduces the equivalent lattice description of the continuous Markovian
process that enables us to calculate anomalous kinetic coefficients in the
vicinity of the boundary. Section~\ref{sec3} is actually the main part of the
paper, it calculates the boundary singularities. The results are used in
Sec.~\ref{sec4} to obtained the desired boundary conditions for the forward and
backward Fokker-Planck equations.

\section{The Chapman-Kolmogorov and Fokker-Planck equations}\label{sec0}

\begin{figure}
\begin{center}
\includegraphics[width=0.9\columnwidth]{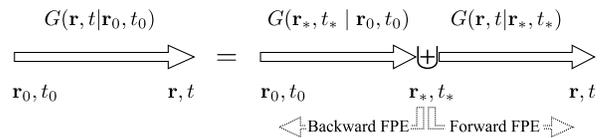}
\caption{Diagram of the Chapman-Kolmogorov equation. The symbol $\uplus$
denotes summation over the intermediate point $\mathbf{r}_*$ and the arrows
illustrate the limit cases $t_*\to t_0+0$ and $t_*\to t-0$ matching the
backward and forward Fokker-Planck equations.} \label{sec0:F1}
\end{center}
\end{figure}

We consider stochastic dynamics of a Markovian system represented as a point
$\mathbf{r}$ belonging to a certain domain $\mathbb{Q}$ in the Euclidean
$M$-dimensional space $\mathbb{R}^M$. The domain $\mathbb{Q}$ is assumed to be
bounded by a smooth hypersurface $\Upsilon$.  When the detailed information
about possible trajectories $\{\mathbf{r}(t)\}$ of the system motion is of
minor importance the conditional probability called also the Green function
$$
    G(\mathbf{r},t|\mathbf{r}_0,t_0) := \mathcal{P}
    \big\{\mathbf{r}_0,t_0 \Rightarrow\mathbf{r},t\big\}
$$
gives us the complete description of system evolution. By definition, the Green
function is the probability density of finding the system at the point
$\mathbf{r}$ at time $t$ provided it was located at the point $\mathbf{r}_0$ at
the initial time $t_0$.

Since Markovian systems have no memory the Green function
$G(\mathbf{r},t|\mathbf{r}_0,t_0)$ obeys the integral Chapman-Kol\-mo\-go\-rov
equation that represents transition of the system from the initial point
$\mathbf{r}_0$ to the terminal one $\mathbf{r}$ within the time interval
$(t_0,t)$ as a complex step via an intermediate point
$\mathbf{r}_*\in\mathbb{Q}$ at a certain fixed moment of time $t_*$ with
succeeding summation over all the possible positions of the intermediate point
(see, e.g., Ref.~\cite{Gardiner})
\begin{equation}\label{sec0:1}
    G(\mathbf{r},t| \mathbf{r}_{0},t_{0})=\iiint\limits_{\mathbb{Q}}
    d\mathbf{r}_{\ast}\,
    G(\mathbf{r},t|\mathbf{r}_{\ast },t_{\ast })
    \,
    G(\mathbf{r}_{\ast},t_{\ast }| \mathbf{r}_{0},t_{0}).
\end{equation}
The time $t_*$ may be chosen arbitrary between the initial and terminal time
moments, $t_*\in [t_0,t]$. Figure~\ref{sec0:F1} visualizes this equation.

Since the domain boundary $\Upsilon$ is considered to be a physical object
special properties will be ascribed to it and itself can affect the system, for
example, trapping it. So the symbol of triple integral is used in
equation~\eqref{sec0:1} to underline this feature and where appropriate it
should be read as
\begin{gather*}
    \iiint\limits_{\mathbb{Q}}d\mathbf{r}\ldots  =
    \int\limits_{\mathbb{Q}^+} d\mathbf{r}\ldots +
    \oint\limits_{\Upsilon} d\mathbf{s}\ldots +
    \oint\limits_{\Upsilon_{tr}} d\mathbf{s}\ldots
\end{gather*}
where the symbol $\mathbb{Q}^+$ denotes the internal points of the domain
$\mathbb{Q}$, the boundary $\Upsilon$ is split from the medium bulk because it
can differ essentially from the medium bulk in properties, and the boundary
traps $\Upsilon_{tr}$ are singled out and treated individually by the same
reasons. To simplify notations a similar rule
\begin{gather*}
    \iint\limits_{\mathbb{Q}} d\mathbf{r}\ldots =
    \int\limits_{\mathbb{Q}^+} d\mathbf{r}\ldots +
    \oint\limits_{\Upsilon} d\mathbf{s}\ldots
\end{gather*}
is also adopted. Such a split of integrals is for treating motion of the system
inside the internal points $\mathbb{Q}^+$, its possible anomalous transport
along the boundary $\Upsilon$, and the trap effect individually. Besides,
according to the probability definition, the equality
\begin{equation}\label{sec0:2}
    \iiint\limits_{\mathbb{Q}}d\mathbf{r}\,
    G(\mathbf{r},t| \mathbf{r}_{0},t_{0}) = 1
\end{equation}
holds when the integration runs over all the possible states of the system
including the boundary  traps $\Upsilon_{tr}$.

In what follows a rather general model for the medium boundary will be studied.
Hear we paid attention only to the fact that the boundary traps have to be
treated individually because the system after been trapped cannot leave the
boundary reaming in a trap forever. As a result if the point $\mathbf{r}_0$
belongs to a trap, then for any internal point $\mathbf{r}$ of the domain
$\mathbb{Q}$ the Green function is equal to zero:
$$
    G(\mathbf{r},t|\mathbf{r}_{0},t_{0})=0\,\quad\text{for}
    \quad \mathbf{r}_0\in \Upsilon_{tr}\,,\;\mathbf{r}\in\mathbb{Q}^+.
$$
Further on the Green function $G(\mathbf{r},t| \mathbf{r}_{0},t_{0})$ for the
\emph{internal} initial and terminal points
$\mathbf{r}_0,\mathbf{r}\in\mathbb{Q}^+$ will be considered. Therefore the
general Chapman-Kolmogorov equation~\eqref{sec0:1} can be reduced by
eliminating the integration over the traps, so becoming
\begin{equation}\label{sec0:1a}
    G(\mathbf{r},t| \mathbf{r}_{0},t_{0})=
    \iint\limits_{\mathbb{Q}}
    d\mathbf{r}_{\ast}\,
    G(\mathbf{r},t|\mathbf{r}_{\ast },t_{\ast })
    \,
    G(\mathbf{r}_{\ast},t_{\ast }| \mathbf{r}_{0},t_{0}).
\end{equation}
In equation~\eqref{sec0:1a} this elimination is pointed out by the absence of
one integral matching the traps, cf. the general formulation~\eqref{sec0:1} of
the Chapman-Kolmogorov equation. Within the given integration rule the equality
matching identity~\eqref{sec0:2} is violated, instead, we have
\begin{equation}\label{sec0:2a}
    \iint\limits_{\mathbb{Q}}d\mathbf{r}\,
    G(\mathbf{r},t| \mathbf{r}_{0},t_{0}) = 1 -
    \oint\limits_{\Upsilon_{tr}} d\mathbf{s}_{tr}
    G(\mathbf{s}_{tr},t| \mathbf{r}_{0},t_{0}) < 1\,,
\end{equation}
where the symbol $\mathbf{s}_{tr}$ stands for the boundary trap located at the
point $\mathbf{s}\in\Upsilon$.

In order to obtain the Fokker-Planck equations two \textit{additional}
assumptions must be adopted. The former is the short time confinement meaning
that on small time scales the system cannot jump over long distances or in
terms of the Green function its first and second moments converge and
\begin{equation}\label{sec0:3}
 \lim_{t\to t_0+0}    \iiint\limits_{\mathbb{Q}}d\mathbf{r}\,
    G(\mathbf{r},t| \mathbf{r}_{0},t_{0})|\mathbf{r}- \mathbf{r}_{0}|^p = 0
    \,,\quad p =1,2\,.
\end{equation}
The latter is the medium local homogeneity. In other words, the medium where
the Markovian process develops, i.e. the domain $\mathbb{Q}$ should be endowed
with characteristics being actually some smooth fields determined inside
$\mathbb{Q}^+$ or at $\Upsilon$ individually. As a result the Green function
$G(\mathbf{r},t|\mathbf{r}_{0},t_{0})$ has to be smooth with respect to all its
arguments for $t>t_0$ and $\mathbf{r}, \mathbf{r}_0\in \mathbb{Q}^+$.

Because the intermediate time $t_*$ entering the Chap\-man-Kolmogorov equation
is any fixed value between the initial and terminal time moments, $t_0<t_*<t$,
there is a freedom to choose it for special reasons.  In particular, the
passage to one of the limits $t_*\to t_0+0$ or $t_*\to t-0$ gives rise to
either the backward or forward Fokker-Planck equation, respectively
(Fig.~\ref{sec0:F1}).

\subsection{The backward Fokker-Planck equation}\label{sec0:sub1}

To implement the limit $t_*\to t_0+0$ let us choose an arbitrary small time
scale $\tau$ and consider the Chapman-Kolmogorov equation for $t_*=t_0+\tau$
and an \emph{internal} point $\mathbf{r}_0$. Then according to the adopted
assumptions the first multiplier $G(\mathbf{r},t|\mathbf{r}_{\ast}, t_{\ast })$
on the right-hand side of ~\eqref{sec0:1a} is a smooth function of both the
argument $\mathbf{r}_*$ and $t_*$ whereas the second one
$G(\mathbf{r}_{\ast},t_{\ast }|\mathbf{r}_{0},t_{0})$ exhibits strong
variations on small spatial scales. So we can expand the function
\begin{equation*}
    G(\mathbf{r},t|\mathbf{r}_0+\mathbf{R},t_0 +\tau)
\end{equation*}
in the Taylor series with respect to the variables $\tau$ and $\mathbf{R = r_*
-r_0}$. The required accuracy is the first order in the time step $\tau$ and
the second order in $\mathbf{R}$ because the characteristic spatial
displacement of the system during time $\tau$ is of order $\tau^{1/2}$. Within
this accuracy it is
\begin{multline}\label{sec0:4}
  G(\mathbf{r},t|\mathbf{r}_0+\mathbf{R},t_0 +\tau)  =
  G(\mathbf{r},t|\mathbf{r}_{0},t_{0})
    \\
  {}+\tau \frac{\partial
  G(\mathbf{r},t| \mathbf{r}_{0},t_{0})}{\partial t_{0}}
  +\sum_{i=1}^{M}R^{i}\nabla _{i}^{0}G(\mathbf{r},t|
  \mathbf{r}_{0},t_0)
  \\
  {}+\frac{1}{2}\sum_{i,j=1}^{M}R^{i}R^{j}\nabla _{i}^{0}
  \nabla_{j}^{0}G(\mathbf{r},t| \mathbf{r}_{0},t_0),
\end{multline}
where the operator $\nabla_i^0 = \partial/\partial x^i_0$ acts only on the
argument $\mathbf{r}_0$ of the Green function. The substitution of
expansion~\eqref{sec0:4} into the Chapman-Kolmogorov equation~\eqref{sec0:1a}
reduces it to the following
\begin{align}
\nonumber
    -\tau
    \frac{\partial G(\mathbf{r},t| \mathbf{r}_{0},t_{0})}{\partial t_{0}}
    & = -\mathfrak{R}(\mathbf{r}_0,t_0,\tau)\,G(\mathbf{r},t|
    \mathbf{r}_{0},t_{0})
\\
\nonumber
    {}& +\sum_{i=1}^{M}\mathfrak{U}^i(\mathbf{r}_0,t_0,\tau)
    \nabla _{i}^{0}G(\mathbf{r},t|\mathbf{r}_{0},t_0)
\\
\label{sec0:5}
    {}&+\sum_{i,j=1}^{M}\mathfrak{L}^{ij}(\mathbf{r}_0,t_0,\tau)
    \nabla_{i}^{0}\nabla_{j}^{0}G(\mathbf{r},t| \mathbf{r}_{0},t_0),
\end{align}
where the quantities
\begin{align}
    \label{sec0:6a}
    \mathfrak{R}(\mathbf{r}_0,t_0,\tau) & = 1 -
    \iint\limits_{\mathbb{Q}} d\mathbf{R} G(\mathbf{r_0+R},t_0+\tau|
    \mathbf{r}_{0},t_{0})\,,
\\
    \label{sec0:6b}
    \mathfrak{U}^i(\mathbf{r}_0,t_0,\tau)&= \iint\limits_{\mathbb{Q}}
    d\mathbf{R}R^i G(\mathbf{r_0+R},t_0+\tau| \mathbf{r}_{0},t_{0})\,,
\\
    \label{sec0:6c}
    \mathfrak{L}^{ij}(\mathbf{r}_0,t_0,\tau)&=\frac12\iint\limits_{\mathbb{Q}}
    d\mathbf{R} R^iR^j G(\mathbf{r_0+R},t_0+\tau| \mathbf{r}_{0},t_{0})
\end{align}
have been introduced. Besides, the first term on the right-hand side of
\eqref{sec0:5} has been assumed to be small and tend to zero as $\tau\to0$
which is justified based on the results to be obtained.

For an internal point $\mathbf{r}_0$ and, thus, separated from the boundary
$\Upsilon$ by finite distance the time step $\tau$ can be chosen so small that
it is possible to construct a neighborhood of the point $\mathbf{r}_0$ with the
following properties. First, deviation of the Green function
$G(\mathbf{r_0+R},t_0+\tau| \mathbf{r}_{0},t_{0})$ from zero outside this
neighborhood is ignorable due the first assumption about the short time
confinement. Second, inside it the medium can be regarded as the homogeneous
space $\mathbb{R}^M$ by virtue of the second assumption on the local
homogeneity. In this case actually replicating the proof of the Law of Large
Numbers using the generation function notion (see, e.g., Ref.~\cite{Gardiner})
it is possible to demonstrate that quantities~\eqref{sec0:6b} and
\eqref{sec0:6c} scale linearly with $\tau$. The difference of
quantity~\eqref{sec0:6a} from zero is ignorable. Therefore for internal points
we can introduce the drift velocity $v^{i}(\mathbf{r},t)$ and the diffusion
tensor $D^{ij}(\mathbf{r},t)$ by the expressions
\begin{align}
    \label{sec0:7b}
    {v}^i(\mathbf{r},t)&= \lim_{\tau \to +0}\frac1\tau
    \int\limits_{\mathbb{Q}^+}
    d\mathbf{R}R^i G(\mathbf{r+R},t+\tau| \mathbf{r},t)\,,
\\
    \label{sec0:7c}
    D^{ij}(\mathbf{r},t)&= \lim_{\tau \to +0}\frac1{2\tau}
    \int\limits_{\mathbb{Q}^+}
    d\mathbf{R} R^iR^j G(\mathbf{r+R},t+\tau| \mathbf{r},t).
\end{align}
Then for the internal points the division of equation~\eqref{sec0:5} by $\tau$
and the succeeding passage to the limit $\tau\to+0$ yield \textit{the backward
Fokker-Planck equation}
\begin{gather}
    \label{BFPE1}
    -\frac{\partial G(\mathbf{r},t| \mathbf{r}_{0},t_{0})}{\partial t_{0}}
     = \bFP \big\{G(\mathbf{r},t| \mathbf{r}_{0},t_{0})\big\}\,,
\\
\intertext{where the backward Fokker-Planck operator is}
    \label{BFPE2}
    \bFP:=
    \sum_{i,j=1}^{M}D^{ij}(\mathbf{r}_0,t_0,\tau)\nabla_{i}^{0}\nabla_{j}^{0}
   +\sum_{i=1}^{M}v^i(\mathbf{r}_0,t_0,\tau)\nabla_{i}^{0}\,.
\end{gather}
We note that the backward Fokker-Planck equation acts on the second spatial
argument of the Green function $G(\mathbf{r},t| \mathbf{r}_{0},t_{0})$.

This Fokker-Planck equation should be supplemented with the initial condition
and the boundary condition. By construction, at the initial time $t_0$ the
system was located at the internal point $\mathbf{r}_0$, so the initial
condition just writes the Green function in the form of the Dirac $\delta
$-function
\begin{equation}\label{FPEinitial}
  \left. G(\mathbf{r},t|\mathbf{r}_{0},t_{0})\right\vert _{t=t_{0}}=
  \delta(\mathbf{r}-\mathbf{r}_{0})\,.
\end{equation}
The boundary condition interrelates the values of the Green function and its
derivatives at the internal points adjacent to the domain boundary $\Upsilon$,
i.e. values obtained by continuation $\mathbf{r}_0\to \mathbf{s}$ from some
internal point $\mathbf{r}_0\in\mathbb{Q}^{M+}$ to a boundary point
$\mathbf{s}\in \Upsilon$.

\subsection{The boundary condition problem for the backward Fokker-Planck equation
and the vector of boundary singularity} \label{sec0:sub}

The direct implementation of the passage to the boundary points, however,
raises a certain problem. Expansion~\eqref{sec0:5} exhibits irregular behavior
within the joint passage to limits $\tau\to+0$ and $\mathbf{r}_0\to
\mathbf{s}$. When the former $\tau\to+0$ precedes the latter $\mathbf{r}_0\to
\mathbf{s}$ no boundary conditions are got at all.

\begin{figure}
\begin{center}
\includegraphics[width=0.8\columnwidth]{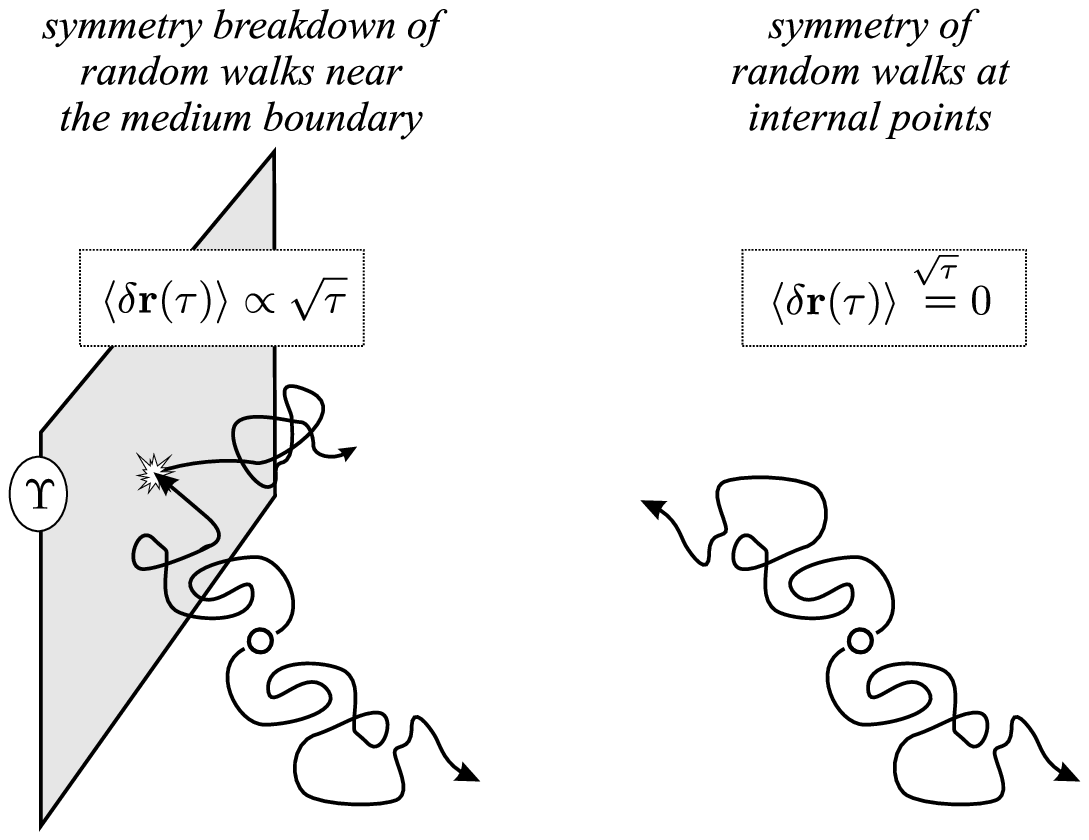}
\caption{The effect of the boundary impermeability on the Markovian system
motion. Schematic illustration.} \label{sec0:F2}
\end{center}
\end{figure}

In the opposite order, i.e. when the passage $\mathbf{r}_0\to \mathbf{s}$ is
performed first, the kinetic coefficients~\eqref{sec0:6a}--\eqref{sec0:6c}
change the scaling type; now they vary with time $\tau$ as $\sqrt{\tau}$ at the
leading order. The matter is that a path of Markovian system is not smooth at
every point and its characteristic variations on small time scales about $\tau$
are proportional to $\sqrt{\tau}$. For the internal points of the domain
$\mathbb{Q}$ the path deviations in opposite directions are equiprobable within
accuracy $\sqrt{\tau}$. As a result the coefficient
$\mathfrak{U}^i(\mathbf{r},t,\tau)$ becomes a linear function of the argument
$\tau$. In some sense the given anomaly in the Markovian dynamics is hidden at
the internal points and reflected only in the linear $\tau$-dependence of the
second order moments $\mathfrak{L}^{ij}(\mathbf{r}_0,t_0,\tau)$ of the Green
function $G(\mathbf{r},t|\mathbf{r}_0,t_0)$. The medium boundary $\Upsilon$
breaks down this symmetry because, in particular, it prevents the system from
getting the points on the opposite side. Since the system displacement remains
the same magnitude the terms $\mathfrak{U}^i(\mathbf{r},t,\tau)$ acquire the
root square dependence on the argument $\tau$. In a certain seance the medium
boundary reveals this anomaly (Fig.~\ref{sec0:F2}). The succeeding division of
expansion~\eqref{sec0:5} by $\tau$ gives rise to singularities of the type
$\tau^{-1/2}$ which will be referred to as \textit{boundary singularities}.

The medium boundary can affect the system dynamics in a more complex way, here,
however, we currently confine our speculations only to the effect of its
impermeability. Since the boundary $\Upsilon$ confines the system motion only
in the normal direction it is quite natural to expect that the boundary
singularities quantified in terms of diverging components
$\mathfrak{U}^i(\mathbf{s},t,\tau)/\tau$ will form a vector object
$\boldsymbol{b}$ that is determined by mutual effect of two factors. The former
is the spatial orientation of the medium boundary $\Upsilon$ described by its
unit normal $\boldsymbol{n}$. The latter is the spatial arrangement and
intensity of random Langevin forces governing stochastic motion of the given
system. They are characterized by the diffusion tensor $D^{ij}(\mathbf{r},t)$.
Within a scalar cofactor we have only one possibility to construct the vector
$\boldsymbol{b} = \{b^i\}$ using the two objects,
\begin{align}
    \label{sec0:bsv1}
    b^i& = \sum_{j=1}^M D^{ij}n^j
    \\
\intertext{or, in the vector form}
    \label{sec0:bsv2}
    \boldsymbol{b}&=\mathbf{D}\cdot\boldsymbol{n}\,.
\end{align}
The validity of this construction will be justified in the present paper and
$\boldsymbol{b}$ will be referred to as the \textit{vector of boundary
singularities}. To be rigorous it should be noted that in the general case the
correct expression for the vector of boundary singularities should use the
operator $D^i_j$ obtained from the diffusion tensor $D^{ij}$ by lowering one of
its indices, namely, $b^i = \sum_j D^i_j n^j$ (such details are discussed in
Sec.~\ref{sec:dtr}). However dealing with orthonormal bases as it is the case
at the initial stage of the current consideration the tensors $D^{ij}$ and
$D^i_j$ coincide with other in the component magnitudes. So in order not to
overload the reader perception and the mathematical constructions expressions
similar to \eqref{sec0:bsv1} will be used where appropriate.

The notion of the boundary singularity vector enables us to write immediately
the desired boundary condition when the medium boundary just confines the
system motion. In this case the first and third terms on the right-hand side of
expansion~\eqref{sec0:5} are absent and the corresponding singularity caused by
the sequence of transitions $\mathbf{r}_0\to\mathbf{s}$ and then $\tau\to0$
takes the form
\begin{multline*}
    \frac1{\sqrt{\tau}} \sum_{i=1}^M b^i(\mathbf{s})\nabla_i^0
    \left. G(\mathbf{r},t|\mathbf{r}_,t_0)\right|_{\mathbf{r}_0\to\mathbf{s}}
\\
    = \frac1{\sqrt{\tau}} \sum_{i,j=1}^M D^{ij}(\mathbf{s},t_0)n^j(\mathbf{s})
    \nabla_i^0
    \left.G(\mathbf{r},t|\mathbf{r}_0,t_0)\right|_{\mathbf{r}_0\to\mathbf{s}}\,.
\end{multline*}
Naturally, for the internal points $\mathbf{r}$ and $\mathbf{r}_0$ the Green
function $G(\mathbf{r},t|\mathbf{r}_0,t_0)$ cannot exhibit any singularity
whence it follows that the cofactor of the singularity $\tau^{-1/2}$ must be
equal to zero, i.e.
\begin{equation*}
     \sum_{i,j=1}^M D^{ij}(\mathbf{s},t_0)n^j(\mathbf{s})
    \nabla_i^0
    \left.G(\mathbf{r},t|\mathbf{r}_0,t_0)\right|_{\mathbf{r}_0\to\mathbf{s}} = 0\,.
\end{equation*}
It is the very know expression for the boundary condition of the backward
Fokker-Planck equation which is typically obtained in another way applying to
physical meaning of the Green function (see, e.g., Ref.~\cite{Gardiner}).

The present paper is devoted to \textit{deriving} the boundary conditions for
the Fokker-Planck equation applying to the notion of the boundary
singularities. A more general situation will be studied justifying also these
qualitative speculations. Currently we can state that the boundary condition
for the backward Fokker-Planck equation should stem from the requirement for
the boundary singularity terms to vanish in expansion~\eqref{sec0:5}, i.e. when
$\mathbf{r}_0\Subset \Upsilon$
\begin{align}
\nonumber
    & -{}^*\mathfrak{R}(\mathbf{r}_0,t_0,\tau)\,G(\mathbf{r},t|
    \mathbf{r}_{0},t_{0})
\\
\nonumber
    {}& +\sum_{i=1}^{M}{}^*\mathfrak{U}^i(\mathbf{r}_0,t_0,\tau)
    \nabla _{i}^{0}G(\mathbf{r},t|\mathbf{r}_{0},t_0)
\\
\label{sec0:5BK}
    {}&+\sum_{i,j=1}^{M}{}^*\mathfrak{L}^{ij}(\mathbf{r}_0,t_0,\tau)
    \nabla_{i}^{0}\nabla_{j}^{0}G(\mathbf{r},t| \mathbf{r}_{0},t_0) = 0\,,
\end{align}
where the symbol $*$ labels the components of the corresponding kinetic
coefficients scaling as $\tau^{1/2}$. It should be pointed out that in
expression~\eqref{sec0:5BK} the argument $\mathbf{r}_0$ is an arbitrary point
of a thing layer $\Upsilon_\tau$ adjacent to the boundary $\Upsilon$, which is
designated by the symbol $\Subset$. When $\tau\to0$ its thickness also tends to
zero (as $\tau^{1/2}$), however, before passing to the limit $\tau\to0$ the
layer $\Upsilon_\tau$ remains volumetric.

Now let us discuss similar problems with respect to the forward Fokker-Planck
equation matching the other possibility of passage to the limit case in the
Chapman-Kolmogorov equation~\eqref{sec0:1a}.

\subsection{The forward Fokker-Planck equation}\label{sec:ffpe}

The Chapman-Kolmogorov equation~\eqref{sec0:1a} also allows for the limit where
the intermediate point tends to the terminal one, i.e. $t_* = t-\tau$ with
$\tau\to+0$. In this case the former cofactor
$G(\mathbf{r},t|\mathbf{r}_*,t-\tau)$ on the right-hand side of \eqref{sec0:1a}
exhibits strong variations on small spatial scales whereas the latter one
$G(\mathbf{r}_*,t-\tau|\mathbf{r}_0,t_0)$ becomes a smooth function of the
argument $\mathbf{r}_*$. Now, however, applying directly to an expansion
similar to that have been used in deriving the backward Fokker-Planck equation
is not appropriate. The matter is that in this way the integration runs over
the initial point $\mathbf{r}_*$ of the Green function
$G(\mathbf{r},t|\mathbf{r}_*,t-\tau)$ and appearing coefficients similar to
quantities~\eqref{sec0:6a}--\eqref{sec0:6c} have another meaning. In
particular, an integral similar to \eqref{sec0:2a} can deviate from unity
essentially.

To overcome this problem the Pontryagin technique is applied \cite{RR}. It is
rather similar to the Kramers-Moyal approach (see, e.g., \cite{Risken}) but is
more suitable for tackling the boundary singularity. Let us consider at the
first step some arbitrary smooth function $\phi(\mathbf{r})$ determined in the
domain $\mathbb{Q}$ and integrate with it both the sides of the
Chapman-Kolmogorov equation~\eqref{sec0:1a}. In this way we get
\begin{multline}\label{sec00:1}
    \iint\limits_{\mathbb{Q}} d\mathbf{r} \phi(\mathbf{r})
    G(\mathbf{r},t_*+\tau| \mathbf{r}_{0},t_{0})
    \\
    {}=
    \iiiint\limits_{\mathbb{Q}\quad\mathbb{Q}}
     d\mathbf{r} d\mathbf{r}_{\ast}\,
     \phi(\mathbf{r})
     \,
    G(\mathbf{r},t_*+\tau|\mathbf{r}_{\ast },t_*)
    \,
    G(\mathbf{r}_{\ast},t_{\ast }| \mathbf{r}_{0},t_{0})\,.
\end{multline}
For a rather small time scale $\tau$ the Green function
$$
G(\mathbf{r},t_*+\tau|\mathbf{r}_{\ast },t_*)
$$
is practically located within some small neighborhood of the point
$\mathbf{r}_*$. Thereby the function $\phi(\mathbf{r})$ can be expanded in the
Taylor series near the point $\mathbf{r}_*$ with respect to the variable
$\mathbf{R=r -r_*}$
\begin{equation*}
  \phi(\mathbf{r})  =  \phi(\mathbf{r}_*)
  +\sum_{i=1}^{M}R^{i}\nabla _{i}^{*}\phi(\mathbf{r}_*)
  +\frac{1}{2}\sum_{i,j=1}^{M}R^{i}R^{j}\nabla _{i}^{*}
  \nabla_{j}^{*}\phi(\mathbf{r}_*)\,.
\end{equation*}
Beside, since the Green function $G(\mathbf{r},t_*+\tau|\mathbf{r}_0,t_0)$
depends smoothly on $\tau$ the expansion
\begin{equation*}
    G(\mathbf{r},t_*+\tau|\mathbf{r}_0,t_0) =
    G(\mathbf{r},t_*|\mathbf{r}_0,t_0)+
    \tau\frac{\partial G(\mathbf{r},t_*|\mathbf{r}_0,t_0)}{\partial t_*}
\end{equation*}
is also justified for a small value of $\tau$.

\begin{figure}
\begin{center}
\includegraphics[width = 0.8\columnwidth]{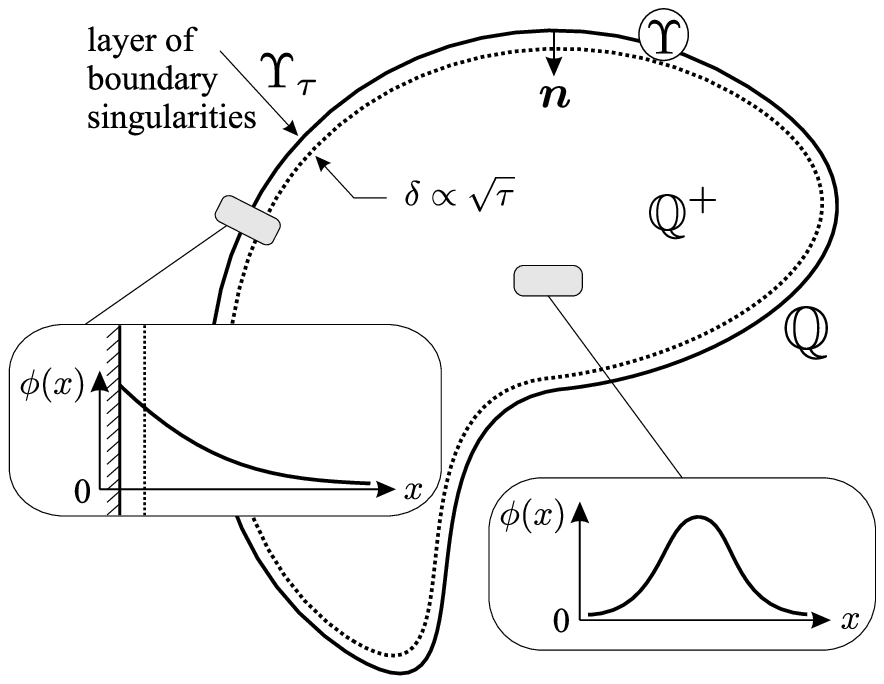}
\caption{Structure of integral~\eqref{sec00:2} and division of the region
$\mathbb{Q}$ into the layer of boundary singularities and the internal points
with regular behavior of the kinetic coefficients.} \label{sec00:F1}
\end{center}
\end{figure}

Then the substitution of the last two expressions into equation~\eqref{sec00:1}
with succeeding integration over $\mathbf{R}$ and the replacement of the dummy
variable $\mathbf{r}_*$ by $\mathbf{r}$ as well as $t_*$ by $t$  yields
{\setlength{\multlinegap}{0pt}
\begin{multline}
\label{sec00:2}
    \iint\limits_{\mathbb{Q}} d\mathbf{r} \phi(\mathbf{r})
    \bigg[
    \tau \frac{\partial G(\mathbf{r},t|\mathbf{r}_0,t_0)}{\partial t}
    \bigg]
\\
\begin{split}
    {}=
    \iint\limits_{\mathbb{Q}}
      d\mathbf{r}
    \Bigg\{
     \phi(\mathbf{r})
     &\bigg[
    -\mathfrak{R}(\mathbf{r},t,\tau)
    \,
    G(\mathbf{r},t| \mathbf{r}_{0},t_{0})
    \bigg]
\\
    {}+ \sum_{i=1}^{M}
    \nabla _{i}\phi(\mathbf{r})
    &\bigg[\mathfrak{U}^{i}(\mathbf{r},t,\tau)
    \,
    G(\mathbf{r},t| \mathbf{r}_{0},t_{0})
    \bigg]
\\
    {}+\sum_{i,j=1}^{M}
    \nabla _{i}\nabla_{j}\phi(\mathbf{r})
    &\bigg[
      \mathfrak{L}^{ij}(\mathbf{r},t,\tau)
    \,
    G(\mathbf{r},t|\mathbf{r}_{0},t_{0})
    \bigg]
    \Bigg\}.
\end{split}
\end{multline}
}%
Here the coefficients $\mathfrak{R}(\mathbf{r},t,\tau)$,
$\mathfrak{U}^{i}(\mathbf{r},t,\tau)$ and
$\mathfrak{L}^{ij}(\mathbf{r},t,\tau)$ again exhibit anomalous behavior within
a narrow layer $\Upsilon_\tau$ adjacent to the medium boundary $\Upsilon$
(Fig.~\ref{sec00:F1}). As should be expected and in accordance with results to
be obtained the thickness of this layer scales with time $\tau$ as
$\tau^{1/2}$. These coefficients themselves also scale as $\tau^{1/2}$. As a
result the corresponding part of integral~\eqref{sec00:2} scales as $\tau$.
Thereby after dividing both of the sides of \eqref{sec00:2} by $\tau$ with the
following passage to the limit $\tau\to0$ the contribution to \eqref{sec00:2}
caused by integration over this layer remains finite. Therefore to analyze the
properties of the integral relation~\eqref{sec00:2} the domain $\mathbb{Q}$ is
split into this layer of boundary singularities and the internal part. After
the passage to the limit $\tau\to0$ this division matches treating individually
the boundary $\Upsilon$ and the internal points $\mathbb{Q}^+$.

Keeping the aforementioned in mind the integral expression~\eqref{sec00:2} is
represented as a sum of two terms, the integral over the layer $\Upsilon_\tau$
denoted with the formal symbol of surface integral and the integral over the
internal part $\mathbb{Q}^+$ of the domain $\mathbb{Q}$
\begin{equation}\label{sec00:3}
    \iint\limits_\mathbb{Q}d\mathbf{r}\ldots =
    \oint\limits_{\Upsilon_\tau} d\mathbf{r}\ldots +
    \int\limits_{\mathbb{Q}^+}d\mathbf{r}\ldots
\end{equation}
Let us consider the second term first. Inside the region $\mathbb{Q}^+$ the
kinetic coefficients $\mathfrak{U}^i(\mathbf{r},t,\tau)$ and
$\mathfrak{L}^{ij}(\mathbf{r},t,\tau)$ behave in regular way, i.e. they scale
as $\tau$ according formulae~\eqref{sec0:7b} and \eqref{sec0:7c}, whereas the
term $\mathfrak{R}(\mathbf{r},t,\tau)$ vanishes at all. So dividing the
corresponding part of the integral relation~\eqref{sec00:2} by $\tau$ and
passing to the limit $\tau\to0$ we have
\begin{multline}\label{sec00:2a}
    \int\limits_{\mathbb{Q}^+} d\mathbf{r} \phi(\mathbf{r})
    \bigg[
    \frac{\partial G(\mathbf{r},t|\mathbf{r}_0,t_0)}{\partial t}
    \bigg]
\\
\begin{split}
    {}= \int\limits_{\mathbb{Q}^+} d\mathbf{r}\Bigg\{\sum_{i=1}^{M}
    \nabla _{i}\phi(\mathbf{r})
    &\bigg[v^{i}(\mathbf{r},t,\tau)
    \,
    G(\mathbf{r},t| \mathbf{r}_{0},t_{0})
    \bigg]
\\
    {}+\sum_{i,j=1}^{M}
    \nabla _{i}\nabla_{j}\phi(\mathbf{r})
    &\bigg[
      D^{ij}(\mathbf{r},t,\tau)
    \,
    G(\mathbf{r},t|\mathbf{r}_{0},t_{0})
    \bigg]
    \Bigg\}.
\end{split}
\end{multline}
Using the Gauss divergence theorem this integral in turn is split into two
parts, surface and volume ones:
\begin{equation}\label{sec00:4}
     \int\limits_{\mathbb{Q}^+} d\mathbf{r}\ldots =
      \oint\limits_{\Upsilon} d\mathbf{s}\ldots +
       \int\limits_{\mathbb{Q}^+} d\mathbf{r}\ldots
\end{equation}
The volume integral has the form
\begin{multline}
\label{sec00:5}
    \int\limits_{\mathbb{Q}^+} d\mathbf{r} \phi(\mathbf{r})
    \bigg[
    \frac{\partial G(\mathbf{r},t|\mathbf{r}_0,t_0)}{\partial t}
    \bigg]
\\
\begin{split}
    {}= \int\limits_{\mathbb{Q}^+} d\mathbf{r}\phi(\mathbf{r})\Bigg\{-\sum_{i=1}^{M}
    \nabla _{i}
    &\bigg[v^{i}(\mathbf{r},t,\tau)
    \,
    G(\mathbf{r},t| \mathbf{r}_{0},t_{0})
    \bigg]
\\
    {}+\sum_{i,j=1}^{M}
    \nabla _{i}\nabla_{j}
    &\bigg[
      D^{ij}(\mathbf{r},t,\tau)
    \,
    G(\mathbf{r},t|\mathbf{r}_{0},t_{0})
    \bigg]
    \Bigg\}.
\end{split}
\end{multline}
The latter equality immediately gives rise to the forward Fokker-Planck
equation.

Indeed, currently $\phi(\mathbf{r})$ is an arbitrary smooth function and no
addition constrains will be imposed further on it for the internal points of
the domain $\mathbb{Q}$. So applying to local variations of $\phi(\mathbf{r})$
at an arbitrary internal point $\mathbf{r}$ (Fig.~\ref{sec00:F1}) we see that
the left and right sides of \eqref{sec00:5} should be equal to each other for
the points $\mathbf{r}\in \mathbb{Q}^+$ individually, getting the forward
Fokker-Planck equation
\begin{gather}
\label{sec00:5fFPE}
    \frac{\partial G(\mathbf{r},t|\mathbf{r}_0,t_0)}{\partial t} =
    \fFP\big\{G(\mathbf{r},t|\mathbf{r}_0,t_0)\big\}
\\
\intertext{with the forward Fokker-Planck operator}
\label{sec00:5fFPEoper}
\begin{split}
    \fFP&\{\lozenge\}   := \sum_{i=1}^{M} \nabla _{i}
\\
    {}&\times\bigg[
    \sum_{j=1}^{M} \nabla_{j}\bigg( D^{ij}(\mathbf{r},t,\tau)\,\lozenge\bigg)
    - v^{i}(\mathbf{r},t,\tau)\,\lozenge \bigg].
\end{split}
\end{gather}
Here the symbol $\lozenge$ stands for a function acted by this operator. It
should be also pointed out that the Fokker-Planck operator acts on the first
spatial argument of the Green function.

The forward Fokker-Planck equation can be also written in the conservation form
\begin{gather}
\label{sec00:5fFPE1}
    \frac{\partial G(\mathbf{r},t|\mathbf{r}_0,t_0)}{\partial t} + \sum_{i=1}^M
    \nabla_i \J^i\big\{G(\mathbf{r},t|\mathbf{r}_0,t_0) \big\} = 0\,,
\\
\intertext{with the probability flux operator $\JJ = \{\J^i\}_{i=1}^M$}
\label{sec00:flux}
    \J^i\{\lozenge\}  :=
    -\sum_{j=1}^{M} \nabla_{j}\bigg( D^{ij}(\mathbf{r},t,\tau)\,\lozenge\bigg)
    + v^{i}(\mathbf{r},t,\tau)\,\lozenge \,.
\end{gather}
The forward Fokker-Planck is naturally supplemented with the same initial
condition~\eqref{FPEinitial}.

\subsection{Boundary relations for the forward Fokker-Planck
equation}\label{sec00:bcffpe}

Splits~\eqref{sec00:3} and \eqref{sec00:4} give rise to two additional terms.
The former one is related to the first split and is the integral over the layer
$\Upsilon_\tau$ of boundary singularities
{\setlength{\multlinegap}{0pt}
\begin{multline}
\label{sec00:10}
    \oint\limits_{\Upsilon_\tau} d\mathbf{r}\,
    G(\mathbf{s},t| \mathbf{r}_{0},t_{0}) \Bigg\{
    \sum_{i=1}^{M}
    \nabla _{i}\phi(\mathbf{s})\, {}^*\mathfrak{U}^{i}(\mathbf{r},t,\tau)
\\
   - \phi(\mathbf{s})\, {}^*\mathfrak{R}(\mathbf{r},t,\tau)
    +\sum_{i,j=1}^{M}
    \nabla _{i}\nabla_{j}\phi(\mathbf{s})\,{}^*\mathfrak{L}^{ij}(\mathbf{r},t,\tau)
    \Bigg\}.
\end{multline}
}%
Here the symbol $d\mathbf{r}$ as well as presence of the argument $\mathbf{r}$
in the singular components of the kinetic coefficients takes into account the
fact that before the passage to the limit $\tau\to0$ the layer $\Upsilon_\tau$
is volumetric. The Green function $G(\mathbf{r},t| \mathbf{r}_{0},t_{0})$ as
well as the test function $\phi(\mathbf{r})$ and its derivatives exhibits minor
variations across the layer  $\Upsilon_\tau$ so their argument $\mathbf{r}$
have been replaced by the corresponding nearest point $\mathbf{s}$ laying on
the boundary $\Upsilon$.

The latter term is due to the part of expression~\eqref{sec00:2a} remaining
after integration using the convergence theorem and can be written in the form
\begin{multline}\label{sec00:2aa}
    \oint\limits_{\Upsilon} d\mathbf{s}
    \sum_{i,j=1}^{M}\nabla_{j}\phi(\mathbf{s})
    n^{i}(\mathbf{s})
    \bigg[
      D^{ij}(\mathbf{s},t,\tau)
    \,
    G(\mathbf{s},t|\mathbf{r}_{0},t_{0})
    \bigg]
\\
    {}= -\oint\limits_{\Upsilon} d\mathbf{s}
    \phi(\mathbf{s})    \sum_{i=1}^{M} n^{i}(\mathbf{s})
    \J^i\big\{G(\mathbf{s},t|\mathbf{r}_{0},t_{0})\}\,,
\end{multline}
where $\boldsymbol{n}(\mathbf{s})= \{n^i(\mathbf{s})\}$ is the unit normal to
the boundary $\Upsilon$ at point $\mathbf{s}$ directed inwards the domain
$\mathbb{Q}$.

Leaping ahead we note that the appropriate choice of the boundary values of the
test function $\phi(\mathbf{s})$ and its derivatives fulfils
equality~\eqref{sec00:10} and, at the next step, gives rise to the required
boundary condition for the forward Fokker-Planck equation. Let us demonstrate
this for the impermeable boundary using the notion of the boundary singularity
vector $\boldsymbol{b}$. Namely, we again assume that for an internal point
$\mathbf{r}$ located in the vicinity of a boundary point $\mathbf{s}$, i.e.
$\mathbf{r}\Subset\mathbf{s}$
\[
    \mathfrak{U}^i(\mathbf{r},t,\tau)\propto b^i(\mathbf{s})
    = \sum_j D^{ij}(\mathbf{s},t)\nu^j(\mathbf{s})\,.
\]
In this case only the first term in equality~\eqref{sec00:10} remains and it is
fulfilled when
\begin{equation}\label{sec00:11}
    \sum_{ij=1}^M D^{ij}(\mathbf{s},t)n^j(\mathbf{s})\nabla_i\phi(\mathbf{s}) = 0\,.
\end{equation}
Equality~\eqref{sec00:11} just relates the boundary values of the test function
$\phi(\mathbf{s})$ with its derivative along the boundary normal
$\boldsymbol{n}(\mathbf{s})$. So for an arbitrary smooth function
$\phi_\Upsilon(\mathbf{s})$ determined at the boundary $\Upsilon$ it is
possible to construct the appropriate function $\phi(\mathbf{r})$ determined in
the domain $\mathbb{Q}$ and meeting equality~\eqref{sec00:11} (see
Fig.~\ref{sec00:F1}). So in the given case the left-hand side and, thus, the
right-hand side of expression~\eqref{sec00:2aa} becomes zero. Since the
integral on the right-hand side of \eqref{sec00:2aa} contains an arbitrary
function $\phi(\mathbf{s})$ determined at the boundary $\Upsilon$ the equality
\begin{equation}\label{sec00:2aaaa}
    \sum_{i=1}^{M} n^{i}(\mathbf{s})
    \J^i\big\{G(\mathbf{s},t|\mathbf{r}_{0},t_{0})\} = 0
\end{equation}
holds for every point of the  boundary $\Upsilon$ individually. This expression
meaning the zero value of the probability flux in the direction normal to the
boundary $\Upsilon$ matches well the physical seance of its impermeability.

However, to \textit{derive} the boundary conditions for the Fokker-Planck
equations more sophisticated constructions are necessary. Besides, in order to
take into account other possible properties of the medium boundary its model
should be specified.

\section{Boundary types}\label{sec1}

\begin{figure}
\begin{center}
\includegraphics[width=\columnwidth]{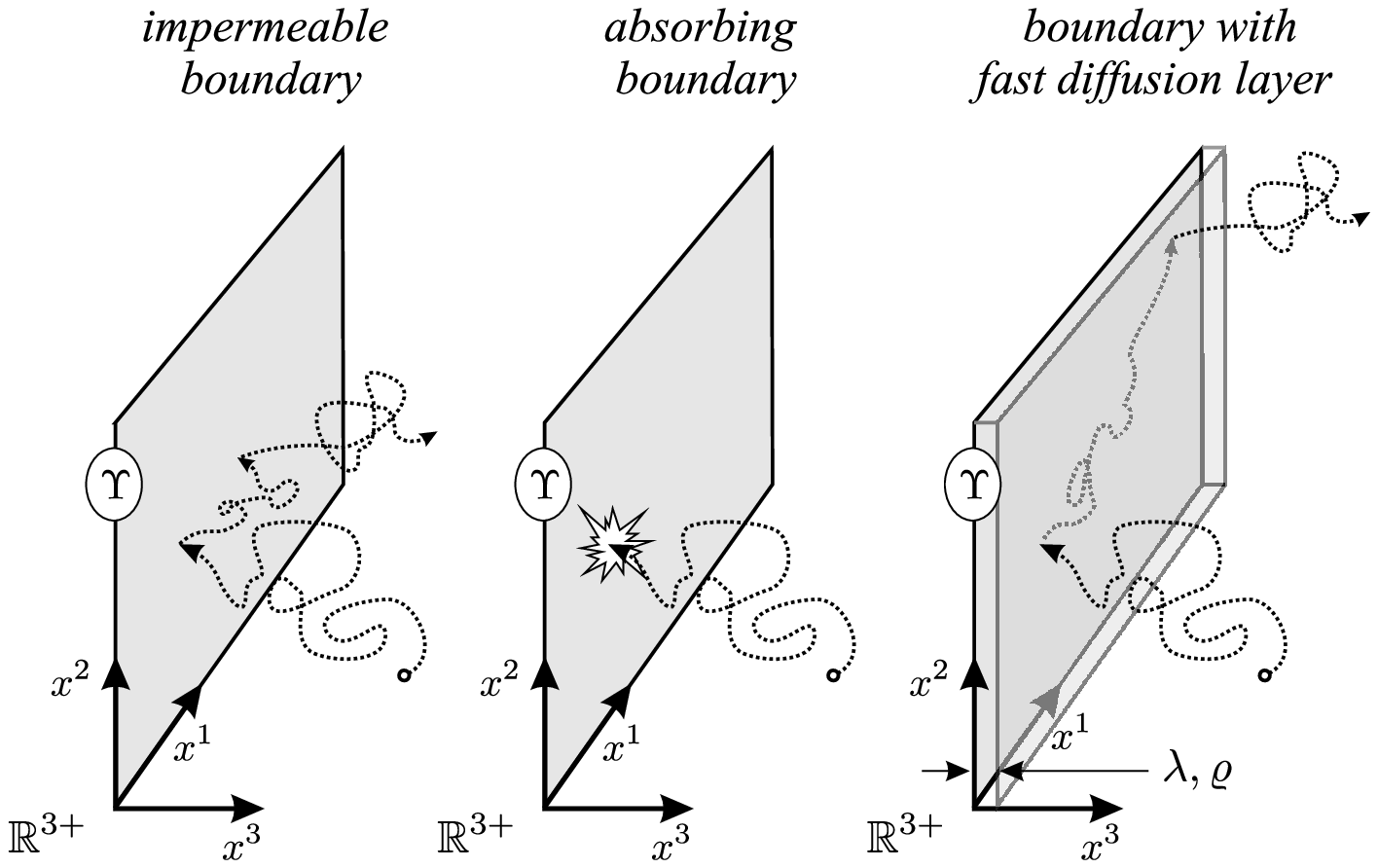}
\end{center}
\caption{Three types of boundaries under consideration.} \label{sec1.F1}
\end{figure}

In the present paper, to be specific, we consider three typical examples of
medium boundaries. They are (\textit{i}) the impermeable boundary,
(\textit{ii}) the boundary absorbing particles, and (\textit{iii}) the boundary
with a thin adjacent layer characterized by extremely high values of the
kinetic coefficients, the fast diffusion boundary (Fig.~\ref{sec1.F1}).

The first type matches a medium whose boundary is similar to its bulk in
properties, the boundary points differ from internal ones only by the absence
of medium points on one side. As a result a random walker hopping over the
medium points just cannot pass through the boundary returning to the medium
bulk after getting it.

The second type is similar to the first one except for the fact that the walker
can be trapped at the boundary and will not return to the medium anymore. In
this case the corresponding boundary conditions are typically used in
describing the first passage time problem or diffusion in solids with fixed
boundary values of impurity concentration $C_s$ (see, e.g.,
Ref.~\cite{Gardiner}). Generally the boundary absorption is described by the
rate $\sigma C_s$, where $\sigma$ is a certain kinetic coefficient.

The third type boundaries are widely met, for example, in polycrystals or
nanoparticle agglomerates. The grain boundaries contain a huge amount of
defects and as a result the diffusion coefficient inside the grain boundaries
can exceed its value in the crystal bulk by many orders. Therefore impurity
propagation in polycrystals is governed mainly by grain boundary diffusion (for
a review see, e.g., Ref.~\cite{GBD} and references therein). In terms of random
walks the effect of the fast diffusion layer is reduced to extremely long
spatial jumps made by an walker inside it. It is natural to characterize such a
boundary layers by its thickness $\lambda$ about the atomic spacing and the
ratio of the diffusion coefficients inside the boundary layer and in the
regular crystal lattice $\varrho\gg1$.

\section{Equivalent lattice representation of random walks near the medium boundary}
\label{sec2}

The derivation of the Fokker-Planck equations, the forward and backward ones,
requires calculation of three quantities $\mathfrak{R}(\mathbf{r},t,\tau)$,
$\mathfrak{U}^i(\mathbf{r},t,\tau)$, and $\mathfrak{L}^{ij}(\mathbf{r},t,\tau)$
specified by expressions~\eqref{sec0:6a}--\eqref{sec0:6c}. They are the moments
of the system displacement $\mathbf{R}$ during the time $\tau$ treated as an
arbitrary small value. In order to obtain the desired boundary conditions these
quantities should be found in the vicinity of the medium boundary $\Upsilon$
or, more precisely, in its neighborhood $\Upsilon_\tau$ of thickness about
$(D\tau)^{1/2}$, where $D$ is the characteristic value of the diffusion tensor
components. To study the boundary effects it suffices to consider a rather
small region wherein the medium and its boundary are practically homogeneous in
properties and, in addition, the boundary geometry is approximated well by some
hyperplane. In this region the system motion will be imitated by random walks
on a lattice constructed as follows.

First, the elementary steps of the random walks on it are characterized by a
time $\tau_a$ such that
\begin{equation}\label{sec2:1}
    \tau_a \ll \tau
\end{equation}
and the arrangement of the lattice nodes, i.e., their spacings $\{a_i\}$ and
the spatial orientation should give us again the same diffusion tensor
$\mathbf{D}$ as well as the drift field $\mathbf{v}$ for the internal points on
time scales $\tau_a\ll t\ll\tau$. The individual hops of a random walker
between the neighboring nodes actually represent a collection of mutually
independent Langevin forces governing the random system motion in the given
continuum. Second, the boundary $\Upsilon$ is represented as a layer of nodes
$\Upsilon_0$ between which the walker can migrate via elementary hops. In other
words, the aforementioned collection of mutually independent Langevin forces
has to contain components acting along the boundary $\Upsilon$ and one
component moving the walker towards or from $\Upsilon$. Other characteristics
of this effective lattice may be chosen for the sake of convenience. At final
stage we should pass to the limit $\tau_a\to0$ returning to the continuous
description.

\subsection{Diffusion tensor representations}\label{sec:dtr}

In order to construct the required lattice let us consider Markovian random
walks $\{\mathbf{r}(t)\}$ in $M$-dimensional Euclidean half-space
$\mathbb{R}^{M+}$ made of vectors
\begin{gather*}
 \mathbf{r} = \{x^1,x^2,\ldots,x^{M}\}\\
 \intertext{such that}
 \mathbf{r}\cdot\boldsymbol{n}:= \sum_{i=1}^M x^i n^i \geq 0\,,
\end{gather*}
where $\boldsymbol{n} = \{n^1,n^2,\ldots,n^M\}$ is a certain unit vector. The
boundary of $\mathbb{R}^{M+}$, i.e. the hyperplane $\Upsilon =
\{\mathbf{r}\cdot\boldsymbol{n} = 0\}$ perpendicular to the vector
$\boldsymbol{n}$ is, in its turn, the Euclidean space $\mathbb{R}^{M-1}$ of
dimension $M-1$. The half-space $\mathbb{R}^{M+}$ and, correspondingly, the
hyperplane $\Upsilon$ are assumed to be homogeneous. The latter means the local
properties of the random walks under consideration to be independent of
position in space; naturally the boundary and internal points are not
equivalent. In particular, the diffusion tensor $\mathbf{D}$ and drift vector
$\mathbf{v}$ are the same at all the internal points of the half-space
$\mathbb{R}^{M+}$.

In this case the components of the drift vector and diffusion tensor are
determined by the expressions (cf. formulae~\eqref{sec0:6a}-\eqref{sec0:6c})
\begin{align}
\label{app2.2}
    v^i & = \frac{1}{\tau}\,\Big<
    \delta X^i(t,\tau)
    \Big>\,,
\\
\label{app2.1}
    {D}{^{ij}} & = \frac{1}{2\tau}\,\Big<
    \left[\delta X^i(t,\tau)-v^i\tau\right]
    \left[\delta X^j(t,\tau)-v^j\tau\right]
    \Big>\,.
\end{align}
Here the random variable $\delta X^i(t,\tau): = x^i(t+\tau)-x^i(t)$ and
$\mathbf{r} = \{x^i\}$ is an arbitrary internal point, the observation time
interval $\tau$ should be chosen to be small enough that the length scale
$(D\tau)^{1/2}$ be much less then the distance between the point $\mathbf{r}$
and the boundary $\Upsilon$, i.e. $D\tau\ll (\mathbf{r}\cdot\boldsymbol{n})^2$,
and the triangular brackets $\langle\ldots\rangle$ stands for averaging over
all the random trajectories passing through the point $\mathbf{r}$ at time $t$.
It should be noted that due to the space homogeneity the passage to the limit
$\tau\to 0$ can be omitted which is necessary in the general case.

In what follows nonorthogonal bases will be used. So, keeping in mind the
tensor notation (see, e.g. Ref.~\cite{tensor1}), the upper and lower indices
will be distinguished. In these terms $\{x^i\}$ or just $x^i$ is a vector,
whereas, the collection of the basis vectors $\mathbf{e}_i$ is a covector.
According to definitions~\eqref{app2.1} and \eqref{app2.2} the objects
${D}{^{ij}}$ and $v^i$ are contravariant tensors. In addition, if the basis
$\mathfrak{e}$ has the form $\mathfrak{e} = \mathfrak{e}_\Upsilon
\oplus\boldsymbol{e}$, where $\mathfrak{e}_\Upsilon$ is the basis of the
hyperplane $\Upsilon$ and the vector $\boldsymbol{e}$ does not lay in it, then
the Greek letters will label the tensor indices corresponding to the hyperplane
$\Upsilon$ to simplify perceiving this fact.

In order to deal with the diffusion tensor in a nonorthogonal basis
$\mathfrak{e} = \{\mathbf{e}_i\}$ the metric tensor is also necessary. It is
defined as
\begin{equation}\label{metric}
{g}{_{ij}}:=(\mathbf{e}_i\cdot\mathbf{e}_j)
\end{equation}
and is the kernel of the scalar product of two vectors $\mathbf{r}$ and
$\mathbf{\bar{r}}$, namely,
\begin{equation*}
    (\mathbf{r}\cdot\mathbf{\bar{r}}):= \sum_{i,j=1}^M
    g_{ij}x^i\bar{x}^j\,.
\end{equation*}
For an orthonormal basis the metric tensor ${g}{_{ij}}= \delta_{ij}$, where
$\delta_{ij}$ is the Kronecker delta. The metric tensor ${g}{_{ij}}$ defines
the conversion of contravariant tensors into covariant ones, in particular,
\begin{gather}
  \label{app2.10}
  {D}{^i_j}  = \sum_{k=1}^M D^{ik} g_{kj}\,, \quad
  {D}{_i^j}  = \sum_{k=1}^M g_{ik}D^{kj}\,,\\
\intertext{as well as}
   \label{app2.11}
   D_{ij}  = \sum_{k,p=1}^M g_{ik}\,g_{jp}\,D^{kp}\,.
\end{gather}
Due to the diffusion tensor ${D}{^{ij}}$ as well as the metric tensor
${g}{_{ij}}$ being symmetric the tensor ${D}{_{ij}}$ is also symmetric, whereas
the tensors $D^i_{\phantom{i}j}$ and $D_j^{\phantom{j}i}$ are identical and so
denoted further as $D^i_j$. The tensor $D^i_j$ can be regarded as a certain
operator $\widehat{\mathcal{D}}$ acting in the space $\mathbb{R}^{M}$ and the
tensor $D_{ij}$ specifies a quadratic form
\begin{equation}\label{app.new1}
    \mathbf{r}\cdot\widehat{\mathcal{D}} \mathbf{r} =
    \sum_{i,j,k=1}^M g_{ij}x^i D^j_k x^k = \sum_{i,j=1}^M D_{ij}x^ix^j\,.
\end{equation}
The quadratic form~\eqref{app.new1} is positive definite. To demonstrate this a
random variable
\begin{align*}
   \delta {L} & = \sum_{p=1}^M
   \big[\delta X^p(t,\tau)- v^p\tau \big]
   (\mathbf{e}_p\cdot\boldsymbol{\ell}\,)\\
   {}&=\sum_{p,i=1}^M\big[\delta X^p(t,\tau)- v^p\tau \big]g_{pi}
   \ell^i
\end{align*}
is considered, where $\boldsymbol{\ell}=\sum_{i=1}^M\mathbf{e}_i \ell^i$ is an
arbitrary vector in the space $\mathbb{R}^{M}$ and the metric tensor
definition~\eqref{metric} has been taken into account. Whence we have a chain
of equalities
\begin{gather*}
    0< \left< \left[\delta{L}\right]^2\right> = \sum_{p,p',i,i'=1}^M
        g_{pi}g_{p'i'} l^i l^{i'}
\\
    {}\times\left<
    \big[\delta X^p(t,\tau)- v^p\tau\big]
    \big[\delta X^{p'}(t,\tau)- v^{p'}\tau\big]
    \right>
\\
    = \sum_{p,p',i,i'=1}^M 2\tau D^{pp'}g_{pi}g_{p'i'} l^i l^{i'}
\\
    {} = \sum_{i,i'=1}^M
    2\tau D_{ii'} l^i l^{i'} = \sum_{p,p',i,i'=1}^M2\tau D^{ii'} l_i l_{i'}.
\end{gather*}
So for any arbitrary vector $l^i$ and covector $l_i$ the inequalities
\begin{align}\label{app.new2}
   \sum_{i,j=1}^M D_{ij}l^il^{j} & > 0, & \sum_{i,j=1}^MD^{ij}l_il_{j} &>0\
\end{align}
hold. The covector and vector representations of the same object are related as
$l_i = \sum_{j=1}^M g_{ij}l^j$; within orthonormal bases they are identical.

Due to the symmetry of the tensor $D_{ij}$ and the quadratic
form~\eqref{app.new1} being positive definite all the eigenvalues of the
operator $\widehat{\mathcal{D}}$ are real positive quantities and its
eigenvectors form a basis in the space $\mathbb{R}^{M}$ which can be chosen to
be orthonormal one, see, e.g., Ref.~\cite{Gantmaher}. In this basis the
diffusion tensor takes the diagonal form. Thereby the corresponding
eigenvectors and eigenvalues specify the directions and intensity of the
mutually independent Langevin forces governing random walks in the medium under
consideration. Unfortunately, in the general case where all the eigenvalues are
nondegenerate this basis is unique. So it cannot be used in constructing the
desired lattice in the vicinity of the medium boundary $\Upsilon$ because one
can meet a situation when none of the basis vectors is parallel to the
hyperplane $\Upsilon$. In order to overcome this problem we will construct a
special nonorthogonal basis applying to the following statement.

\begin{proposition}\label{app2.Prep2}

Let $\mathbb{R}^{M+}=\{\mathbf{r}\cdot\boldsymbol{n}>0\}$ be a homogeneous
half-space bounded by the hyperplane $\Upsilon =
\{\mathbf{r}\cdot\boldsymbol{n}=0\}$ and $\mathfrak{e} =
\{\mathbf{e}_1,\mathbf{e}_2,\ldots\mathbf{e}_M\}$ be a fixed arbitrary basis of
$\mathbb{R}^M$. In this basis the components of the diffusion tensor
$\{D^{ij}\}$ as well as the matric tensor $\{g_{ij}\}$ are given. Then there is
a basis $\mathfrak{b} = \mathfrak{b}_\Upsilon\oplus\boldsymbol{b}_M$ with the
following properties.

First, it is composed of a certain orthonormal basis $\mathfrak{b}_\Upsilon$ of
the hyperplane $\Upsilon$ and a unit vector $\boldsymbol{b}_M$ not belonging to
$\Upsilon$ that is determined by the expression
\begin{equation}\label{app2.Prep2.D1}
    \boldsymbol{b}_M = \frac1\omega\sum_{i,j=1}^M\mathbf{e}_i D^i_jn^j\,.
\end{equation}
Here according to the construction of the half-space $\mathbb{R}^{M+}$
$\boldsymbol{n} = \{n^1,n^2,\ldots n^M\}$ is the unit vector normal to the
hyperplane $\Upsilon$ and the normalization factor
\begin{equation}
 \label{app2.Prep2.D2}
 \omega =  \left[\sum_{i,j,p,k=1}^M g_{ij}D^i_pD^j_{k}n^pn^k \right]^{1/2}.
\end{equation}
Second, in the basis $\mathfrak{b}$ the diffusion tensor takes the diagonal
form
\begin{equation}\label{app2.Prep2.D3}
 \|D^{ij}\| =  \left\|%
\begin{array}{cccc}
  \mathcal{D}_1 &     0     & \ldots & 0        \\
     0      & \mathcal{D}_2 & \ldots & 0        \\
  \vdots    & \vdots    & \ddots & \vdots   \\
    0       &    0      & \ldots & \mathcal{D}_M\\
\end{array}%
\right\|\,,
\end{equation}
where all its diagonal components are positive quantities, $\{\mathcal{D}_i
>0\}$, with the value $\mathcal{D}_M$ being given by the expression
\begin{equation}\label{app2.Prep2.D4}
 \mathcal{D}_M  = \omega^2 \left[\sum_{i,j=1}^M
    D_{ij}n^in^j\right]^{-1}\,.
\end{equation}

Third, let, in addition, the initial basis be of the form
$\mathfrak{e}=\mathfrak{e}_\Upsilon\oplus\,\boldsymbol{n}$, where
$\mathfrak{e}_\Upsilon = \{\mathbf{e}_\alpha\}_1^{M-1}$ is a certain basis of
the hyperplane $\Upsilon$, and $\Hat{U}_\Upsilon =
\|u^\alpha_{\phantom{\alpha}\beta}\|$ be the transformation of the hyperplane
$\Upsilon$ mapping  the basis $\mathfrak{b}_\Upsilon$ onto the basis
$\mathfrak{e}_\Upsilon$, i.e.,
$\mathfrak{b}_\Upsilon\overset{\Hat{u}}\mapsto\mathfrak{e}_\Upsilon$. By
mapping $\boldsymbol{b}_M\mapsto \boldsymbol{n}$ the transformation
$\Hat{U}_\Upsilon$ is complemented to a certain transformation $\Hat{U}$ of the
compleat space $\mathbb{R}^M$, namely, if $\mathbf{r}$ is an arbitrary vector
of the space $\mathbb{R}^M$ with the coordinates specified by its expansion
over the bases $\mathfrak{e}$ and $\mathfrak{b}$:
\begin{equation}\label{add.13}
    \mathbf{r} =\sum_{\gamma=1}^{M-1} \mathbf{e}_\gamma x^\gamma + \boldsymbol{n} x^M \equiv
    \sum_{\gamma=1}^{M-1} \mathbf{b}_\gamma \zeta^\gamma + \boldsymbol{b}_M\zeta^M,
\end{equation}
then its coordinates are related by the expressions
\begin{subequations}\label{app2.Prep2.D5}
\begin{align}
 \label{app2.Prep2.D5a}
  \zeta^\alpha & = \sum_{\gamma=1}^{M-1} u^\alpha_{\ \gamma}
  \Big(x^\gamma - \frac1{D^{MM}}D^{\gamma M}\,x^M\Big)\,,
  \\
  \label{app2.Prep2.D5aM}
  \zeta^M & = \frac{\omega}{D^{MM}}\, x^M\,,
  \\
  \intertext{and for the inverse transformation}
 \label{app2.Prep2.D5b}
  x^\alpha & = \sum_{\gamma=1}^{M-1}\breve{u}^\alpha_{\phantom{\alpha} \gamma}\,
  \zeta^\gamma + \frac1\omega D^{\alpha M}\,\zeta^M\,,
  \\
   \label{app2.Prep2.D5bM}
  x^M  &= \frac{D^{MM}}{\omega}\,\zeta^M\,.
\end{align}
\end{subequations}
Here $\Hat{U}_\upsilon^{-1} = \|\breve{u}^\alpha_{\ \beta}\|$ is the operator
inverse to the operator $\Hat{U}_\Upsilon$, i.e. meeting the identity
$\sum_{\gamma=1}^{M-1}\breve{u}^\alpha_{\ \gamma}u^\gamma_{\
\beta}=\delta^\alpha_\beta$. Besides, the equality
\begin{equation}\label{app2.Prep2.D6}
 \sum_{\gamma=1}^{M-1}
 \breve{u}^\alpha_{\ \gamma} \breve{u}^\beta_{\ \gamma}\mathcal{D}_\gamma =
 D^{\alpha\beta}-\frac1{D^{MM}}D^{\alpha M}D^{\beta M}
\end{equation}
holds.
\end{proposition}
Since the proof of this proposition requires just formal mathematical
manipulations related weakly to the subject matter of the paper it is presented
in the individual Appendix~\ref{appendix}.
\vspace*{0.5\baselineskip}

\noindent \textit{Comments on Proposition~\ref{app2.Prep2}\;}
First, it is worthwhile to note that the basis vector $\boldsymbol{b}_M$
constructed by expression~\eqref{app2.Prep2.D1} is actually the vector
$\boldsymbol{b}$ of boundary singularities (expression~\eqref{sec0:bsv1})
normalized to unity.

Second, for the initial basis $\mathfrak{e}$ of the general form actually
expression~\eqref{app2.Prep2.D6} persuades us to introduce the surface
diffusion tensor
\begin{equation}\label{IamG1}
    \mathfrak{D}^{ij}:= D^{ij} -
    \frac{\sum_{p,k=1}^{M} D^i_pD^j_kn^p n^p}{\sum_{p,k=1}^M D_{pk}n^k n^p}
\end{equation}
that describes the system random motion along the hyperplane $\Upsilon$.
Indeed, in a basis $\mathfrak{b}_\Upsilon\oplus\boldsymbol{n}$ the components
of this tensor belonging to the hyperplane $\Upsilon$ coincide with ones given
by expression~\eqref{app2.Prep2.D6} and are equal to zero when one of its
indices matches the vector $\boldsymbol{n}$.

Third, when the initial basis $\mathfrak{e}$ is orthonormal the expressions of
Proposition~\ref{app2.Prep2} can be simplified. Indeed, in this case the matric
tensor $g_{ij}:=(\mathbf{e}_i\cdot\mathbf{e}_j) = \delta_{ij}$ is the unit
matrix and it is possible not to distinguish between the upper and lower tensor
indices, in particular, all the components $D_{ij} = D^i_j = D^{ij}$ are
identical. If, in addition, the initial basis has the form
$\mathfrak{e}=\mathfrak{e}_\Upsilon\oplus\,\boldsymbol{n}$
expressions~\eqref{app2.Prep2.D1}--\eqref{app2.Prep2.D4} become
\begin{align}
    \label{app2.Comm.1}
    \boldsymbol{b}_M & = \frac1\omega\,\left[\sum_{\gamma = 1}^{M-1} \mathbf{e}_\gamma D_{\gamma M}
    +\boldsymbol{n} D_{MM} \right]\,,\\
\intertext{where the coefficient}
    \label{app2.Comm.2}
    \omega  & = \left[\sum_{\gamma = 1}^{M-1} D_{\gamma M}^2+ D_{MM}^2
    \right]^{1/2}\\
\intertext{and the value}
    \label{app2.Comm.3}
    \mathcal{D}_M & = \frac1{D_{MM}}\sum_{\gamma = 1}^{M-1} D_{\gamma M}^2+
    D_{MM}\,.
\end{align}
Besides, the inverse transformation matrix $\|\breve{u}_{\alpha\beta}\|$
coincides with the direct transformation matrix transposed, i.e.
$\breve{u}_{\alpha\beta}= u_{\beta\alpha}$. $\Box$
\vspace*{0.5\baselineskip}

Proposition~\ref{app2.Prep2} prompts us to use the basis $\mathfrak{b}=
\{\mathbf{b}_i\}$ in describing random walks in the half-space
$\mathbb{R}^{M+}$. For its \textit{internal} points the continuous random walks
are represented as a collection of mutually independent one-dimensional
Markovian processes $\{\zeta^i(t)\}$
\begin{equation}\label{app2.30}
    \mathbf{r}(t) = \mathbf{b}_i \zeta^i(t) =  \mathbf{b}_i \int^t_0 dt'\,\xi^{i}(t')\,,
\end{equation}
where the Langevin random forces $\{\xi^i(t)\}$ meet the correlations
\begin{align}
        \label{app2.30cor1}
  \Big< \xi^{i}(t)\Big > & =  v^i\,,\\
        \label{app2.30cor2}
  \Big< \xi^{i}(t) \xi^{i'}(t') \Big > & = 2\mathcal{D}_i
  \delta_{ii'}\delta(t-t')\,,
\end{align}
and $\{v^i\}$ are the components of the drift velocity
$\mathbf{v}=\mathbf{b}_iv^i$ in the basis $\mathfrak{b}$. As could be shown
directly these random forces lead to expressions~\eqref{app2.2} and
\eqref{app2.1}.

\subsection{Equivalent lattice random walks}\label{sec:elrw}

\begin{figure}
\begin{center}
\includegraphics[width = 0.8\columnwidth]{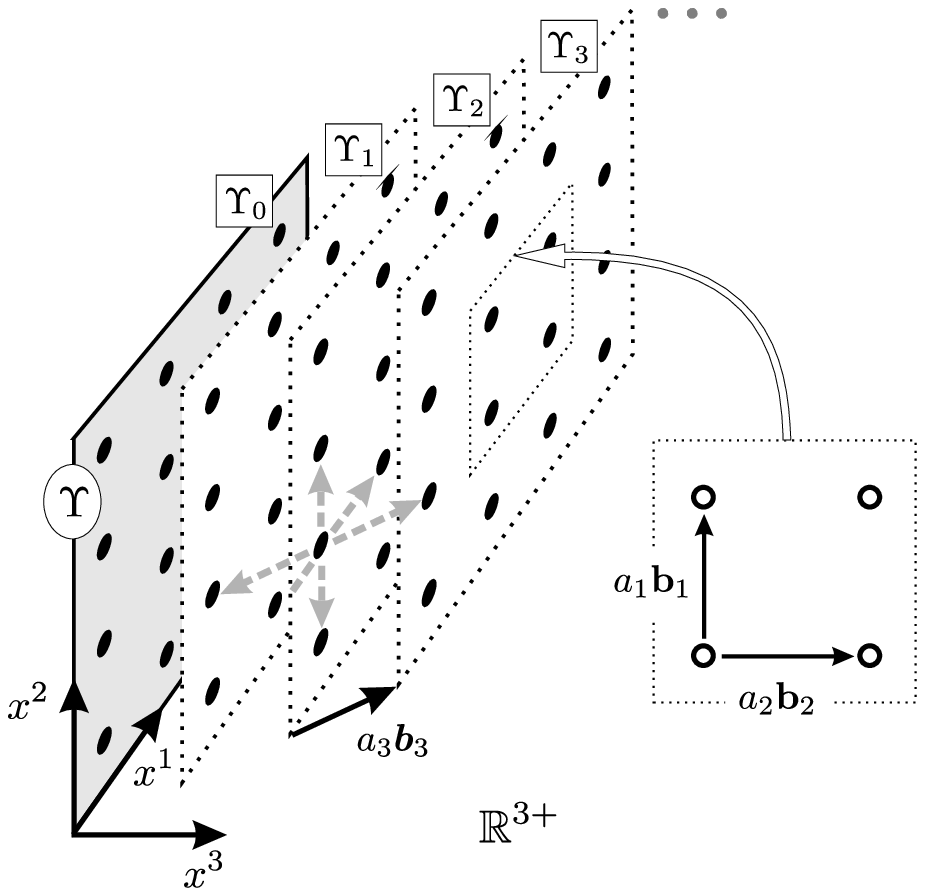}
\end{center}
\caption{The lattice random walks imitating continuous Markovian process in the
half-space $R^{3+}$. Here $\Upsilon$ is the boundary of $R^{3+}$, the axes
$x^1$, $x^2$ are chosen to be directed along the vectors $\mathbf{b}_1$,
$\mathbf{b}_2$ of the basis $\mathfrak{b}_\Upsilon$, the axis $x^3$ is normal
to the plane $\Upsilon$, whereas the basic vector $\boldsymbol{b}_3$ is not
normal to it in the general case. The values $a_1,a_2,a_3$ are the lattice
spacings and grey arrows show possible hopes to the nearest neighbors.}
\label{sec2:F1}
\end{figure}

The desired lattice is constructed as follows (see also Fig.~\ref{sec2:F1} for
illustration). At the first step a set of nodes $\{\mathbf{a}_\Upsilon \}$ is
fixed on the boundary $\Upsilon$ such that
\begin{equation}\label{app2.31}
  \mathbf{a}_\Upsilon(\mathbf{n}_\Upsilon) = \sum^{M-1}_{\alpha=1}
            \mathbf{b}_\alpha a_\alpha n_\alpha \,,
\end{equation}
where $\mathbf{n}_\Upsilon= \{n_1,n_2,\ldots,n_{M-1}\}$ is a collection of
integers taking values in $\mathbb{Z}$ and the lattice spacings $a_\alpha$ are
chosen to be equal to
\begin{subequations}\label{app2.32}
\begin{equation}\label{app2.32a}
    a_\alpha = \sqrt{2\tau_a M \mathcal{D}_\alpha}\,.
\end{equation}
Here $\tau_a$ is any small time scale meeting inequality~\eqref{sec2:1} and
being the time step of lattice random works; an walker hops to one of the
nearest neighbors in time $\tau_a$. Such jumps are illustrated by grey arrows
in Fig.~\ref{sec2:F1}. These nodes are regarded as the boundary layer
$\Upsilon_0$ of the lattice to be constructed. Then the layer $\Upsilon_0$ as a
whole is shifted inwards the region $\mathbb{R}^{M+}$ by the vector
$a_M\boldsymbol{b}_M$, where
\begin{equation}\label{app2.32b}
    a_M = \sqrt{2\tau_a M \mathcal{D}_M}\,.
\end{equation}
\end{subequations}
Then this new layer $\Upsilon_1$ in turn is shifted by the same vector
$\mathbb{R}^{M+}$, giving rise to the next layer $\Upsilon_2$ of nodes and so
on. In this way we construct the system of layers $\{\Upsilon_k\}$ making up
the desired lattice and exactly random walks on this lattice will imitate the
continuous process in the half-space $\mathbb{R}^{M+}$.

Let us now specify the probability of hops from an internal node $\mathbf{n}$
to one of its nearest neighbors $\mathbf{n'}$ along a basis vector
$\mathbf{b}_i$ by the expression
\begin{equation}\label{app2.33}
    P_{\mathbf{n}\mathbf{n}'} = \frac1{2M} +
    \frac{\tau_a}{2a_i}v^{i}\chi_i\,.
\end{equation}
Here the random value $\chi_i = \pm 1$ takes into accounts the possibility of
jumps along the vector $\mathbf{b}_i$ or in the opposite direction. The
sequence of such hops with time step $\tau_a$ represents equivalently the
continuous process rather far from the boundary $\Upsilon$. Indeed, due to the
law of large numbers (see, e.g. \cite{FellerII}) two Markovian processes are
identical if on a rather small time scale both of them lead to the same mean
and mean-square values of the system displacement. By virtue of \eqref{app2.33}
one hop of the walker is characterized by the following mean values of its
displacement $\delta\mathbf{r} = \mathbf{b}_i\delta\zeta^i$
\begin{align}
    \label{app2.eq1a}
    \sum_{\mathbf{n}'} P_{\mathbf{n}\mathbf{n}'}\,\delta\zeta^i_{\mathbf{n}\mathbf{n'}}
    & = \tau_a v^i\,,
    \\
    \label{app2.eq1b}
    \sum_{\mathbf{n}'} P_{\mathbf{n}\mathbf{n}'}
    \,\delta\zeta^i_{\mathbf{n}\mathbf{n'}}
    \,\delta\zeta^j_{\mathbf{n}\mathbf{n'}} & =
    2\tau_a\mathcal{D}_i\delta_{ij}\,,
\end{align}
where the sums run over all the nearest neighbors $\mathbf{n'}$ of the node
$\mathbf{n}$. According to \eqref{app2.2} and \eqref{app2.1} actually the same
mean values of the system displacement during the time interval $\tau_a$ are
given by the continuous random process. Rigorously speaking, the latter mean
value and one corresponding to the continuous random process are not identical,
but their difference
\begin{equation*}
    (\mathbf{b}_i\cdot\mathbf{b}_j) v^iv^j\tau_a^2
\end{equation*}
is of the second order in the time scale $\tau_a$ whereas the leading terms are
of the first order. Thereby choosing the time scale $\tau_a$ to be sufficiently
small we can make this difference ignorable.

\subsection{Properties of the boundary layer $\Upsilon_0$}

In order to describe the boundary effects on random walks special properties
should be ascribed to the nodes of the boundary layer $\Upsilon_0$. It is
worthwhile to noted that it is the place where the model of the medium boundary
does appear for the first time.

\begin{figure}
\begin{center}
\includegraphics[width=0.9\columnwidth]{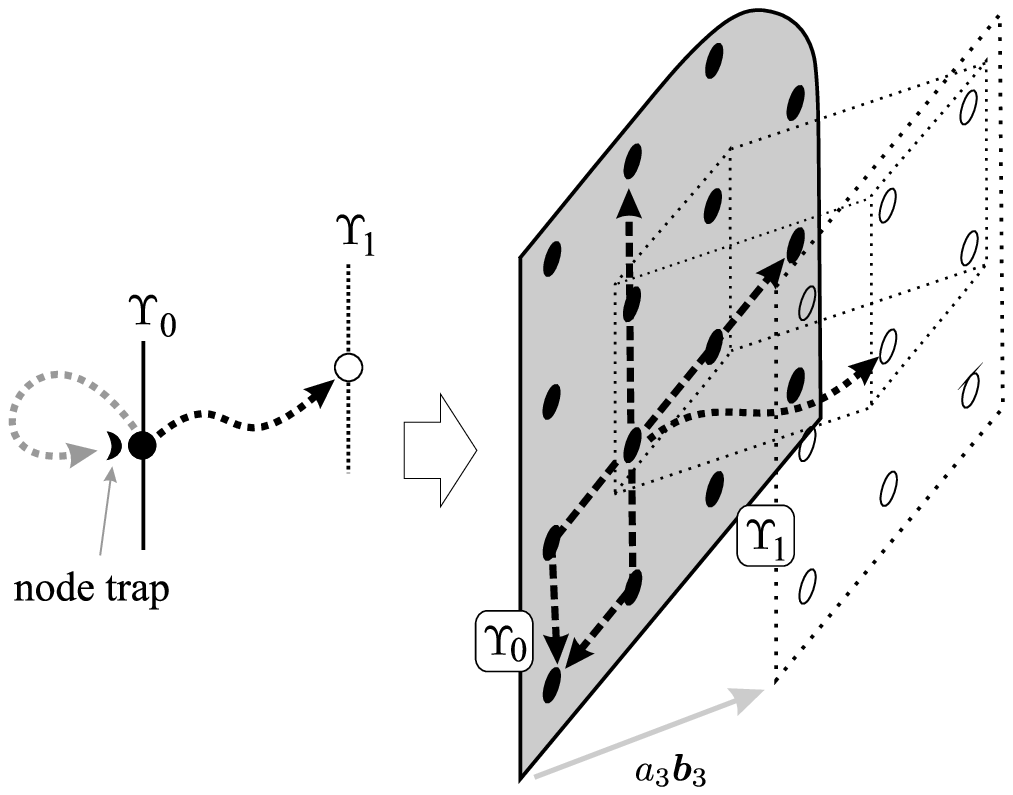}
\caption{Characteristic properties of random walks in the boundary layer
$\Upsilon_0$. The left inset visualizes possible hops from the boundary layer.
The main fragment illustrates the walker jumps inside the boundary layer
$\Upsilon_0$ which can be complex and comprise many elementary hops. The latter
feature imitates possible fast diffusion inside a certain thin layer adjacent
to crystal boundaries} \label{sec2:F2}
\end{center}
\end{figure}

Keeping in mind the boundary types discussed in Sec.~\ref{sec1}, first, each
boundary node is regarded as a unit of two elements, the lattice node itself
and a trap. If a walker jumps to a trap it will never return to the lattice
nodes. The introduction of traps mimics the absorption effect of medium
boundaries. Second, possible fast diffusion inside a thin layer adjacent to
medium boundaries is imitated in terms of multiple steps over the boundary
nodes during the time interval $\tau_a$. These constructions are illustrated in
Fig.~\ref{sec2:F2}.

For the walker located at a certain boundary node the probabilities of hopping
to the internal neighboring node, $P_l$, or being trapped, $P_{tr}$ are
specified as
\begin{align}
    \label{app2.34a}
    P_l  & = \frac{1-\sigma_a}{M}\,, &
    P_{tr} & = \frac{\sigma_a}{M}\,,
\end{align}
where the coefficient $\sigma_a$ quantifies the trapping (absorption) effect.
Leaping ahead, we note that the coefficient $\sigma_a$ can be assumed to be a
small value because its magnitude $\sigma_a\to 0$ as $\tau_a\to0$ within the
collection of lattices leading to the equivalent description of the random
walks on time scales $\tau_a\ll t\ll\tau$. These probabilities have been chosen
to constitute the probability of walker motion along the direction of the basis
vector $\boldsymbol{b}_M$ equal to the same value for the internal points,
\begin{equation*}
    P_l+P_{tr} = \frac1M\,.
\end{equation*}

Therefore the probability for the walker being initially at a boundary node
$\mathbf{n}_\Upsilon$ to make a jump within the boundary layer $\Upsilon_0$ is
\begin{equation}\label{app.new10}
    P_\Upsilon = \frac{M-1}{M}\,.
\end{equation}
At first, let us consider the case where such jumps are the elementary hops to
one $\mathbf{n}'_\Upsilon$ of the nearest neighboring nodes in $\Upsilon_0$.
Then, following actually construction~\eqref{app2.33} its conditional
probability is written as
\begin{equation}\label{app2.35}
    P^{(1)}_{\mathbf{n}^{\vphantom{\prime}}_\Upsilon\mathbf{n}'_\Upsilon} =
    \frac1{2(M-1)} +
    \frac{M}{(M-1)}\,
    \frac{\tau_a}{2a_\alpha}\,v_\Upsilon^\alpha\chi_\alpha\,.
\end{equation}
Here, as before, the value $\chi_\alpha =\pm 1$ is ascribed to the walker hop
along the basis vector $\mathbf{b}_\alpha$ or in the opposite direction,
$v_\Upsilon^\alpha$ are the components of the drift velocity inside the
boundary layer in the basis $\mathfrak{b}_\Upsilon$. It should be noted that
regular drift inside the boundary layer and the medium can be different in
nature, which is allowed for by the index $\Upsilon$ at the boundary components
of the drift velocity. The adopted expression~\eqref{app2.35}, as it must,
obeys equalities similar to expressions~\eqref{app2.eq1a},  \eqref{app2.eq1b},
namely, for the displacement
$\delta\mathbf{r}^{\phantom{\prime}}_\Upsilon=\mathbf{b}_\alpha\delta\zeta^\alpha$
along the boundary $\Upsilon$
\begin{align}
    \label{app2.eq2a}
    P_\Upsilon
    \sum_{\mathbf{m} \in\Upsilon}
    P^{(1)}_{\mathbf{n}^{\vphantom{\prime}}_\Upsilon\mathbf{m}}\,
    \delta\zeta^\alpha_{\mathbf{n}^{\vphantom{\prime}}_\Upsilon\mathbf{m}}
    & = \tau_a v^\alpha_\Upsilon\,,
    \\
    \label{app2.eq2b}
    P_\Upsilon
    \sum_{\mathbf{m}\in\Upsilon}  P^{(1)}_{\mathbf{n}^{\vphantom{\prime}}_\Upsilon\mathbf{m}}
    \,\delta\zeta^\alpha_{\mathbf{n}^{\vphantom{\prime}}_\Upsilon\mathbf{m}}
    \,\delta\zeta^\beta_{\mathbf{n}^{\vphantom{\prime}}_\Upsilon\mathbf{m}} & =
    2\tau_a\mathcal{D}_\alpha\delta_{\alpha\beta}\,.
\end{align}
The fast diffusion inside the boundary layer $\Upsilon$ is imitated by complex
jumps made up of $g$ successive elementary hops within the time $\tau_a$. In
this case the walker can get not only the nearest neighboring nodes but also
relatively distant ones. The conditional probability of such a $g$-fold jump
from node $\mathbf{n}^{\vphantom{\prime}}_\Upsilon$ to node
$\mathbf{n}'_\Upsilon$  is given by the expression
\begin{multline}\label{app2.36}
    P^{(g)}_{\mathbf{n}^{\vphantom{\prime}}_\Upsilon\mathbf{n}'_\Upsilon} =
    \sum_{\mathbf{m}_1,\mathbf{m}_2,\ldots,\mathbf{m}_{g-1}\in \Upsilon}
    P^{(1)}_{\mathbf{n}^{\vphantom{\prime}}_\Upsilon\mathbf{m}_1^{\vphantom{\prime}}}\times
    P^{(1)}_{\mathbf{m}_1^{\vphantom{\prime}}\mathbf{m}_2^{\vphantom{\prime}}}
    \\
    \cdots\times
    P^{(1)}_{\mathbf{m}^{\vphantom{\prime}}_{g-2}\mathbf{m}_{g-1}^{\vphantom{\prime}}}
    \times
    P^{(1)}_{\mathbf{m}^{\vphantom{\prime}}_{g-1}\mathbf{n}'_\Upsilon}\,.
\end{multline}
By virtue of \eqref{app2.eq2a} and \eqref{app2.eq2b} the probability function
$P^{(g)}_{\mathbf{n}^{\vphantom{\prime}}_\Upsilon\mathbf{n}'_\Upsilon}$ of
$g$-fold jumps gives the following values for the first and second moments of
the walker displacement $\delta\mathbf{r}^{\phantom{\prime}}_\Upsilon =
\sum_{\alpha = 1}^{M-1}\mathbf{b}_\alpha\delta\zeta^\alpha$ in the layer
$\Upsilon_0$
\begin{align}
    \label{app2.eq3a}
    P_\Upsilon
    \sum_{\mathbf{m} \in\Upsilon}
    P^{(g)}_{\mathbf{n}^{\vphantom{\prime}}_\Upsilon\mathbf{m}}\,
    \delta\zeta^\alpha_{\mathbf{n}^{\vphantom{\prime}}_\Upsilon\mathbf{m}}
    & = g\tau_a v^\alpha_\Upsilon\,,
    \\
    \nonumber
    P_\Upsilon
    \sum_{\mathbf{m}\in\Upsilon}  P^{(g)}_{\mathbf{n}^{\vphantom{\prime}}_\Upsilon\mathbf{m}}
    \,\delta\zeta^\alpha_{\mathbf{n}^{\vphantom{\prime}}_\Upsilon\mathbf{m}}
    \,\delta\zeta^\beta_{\mathbf{n}^{\vphantom{\prime}}_\Upsilon\mathbf{m}} & =
    2(g\tau_a)\mathcal{D}_\alpha\delta_{\alpha\beta}
    \\
    \label{app2.eq3b}
    {}& + (g\tau_a)^2 \frac{M}{(M-1)} v^\alpha_\Upsilon v^\beta_\Upsilon
    \,.
\end{align}
In expression~\eqref{app2.eq3b} we again have ignored terms of order
$g\tau^2_a$ because the displacement of a walker along the boundary $\Upsilon$
caused by its migration inside the layer $\Upsilon_0$ is considerable only for
$g\gg1$ as will be seen further. In latter case the conditional
probability~\eqref{app2.36} of transition from the node
$\mathbf{n}^{\vphantom{\prime}}_\Upsilon$ to the node
$$
   \mathbf{n}'_\Upsilon = \mathbf{n}_\Upsilon + \sum_\alpha
   \mathbf{b}_\alpha m_\alpha
   \quad (\text{$\{m_\alpha\}$ are integers})
$$
can be approximated by the Gaussian distribution
\begin{multline}\label{app2.36add1}
 P^{(g)}_{\mathbf{n}^{\vphantom{\prime}}_\Upsilon\mathbf{n}'_\Upsilon} =
 \Big(\frac{M-1}{2\pi g} \Big)^{\tfrac{M-1}2}
 \\\times
 \exp\Bigg\{-\frac{(M-1)}{2g}\sum_{\alpha=1}^{M-1}
 \Big[m_\alpha - \frac{gM}{(M-1)}\frac{\tau_a v_\Upsilon^\alpha}{a_\alpha}\Big]^2
 \Bigg\}
\end{multline}
by virtue of the law of large numbers and expressions~\eqref{app2.32a},
\eqref{app2.eq3a}, \eqref{app2.eq3b}.

The desired lattice random walks imitating the continuous Markovian process in
the vicinity of the medium boundary $\Upsilon$ is constructed.

\section{Boundary singularities}\label{sec3}

As discussed in Sec.~\ref{sec0:sub} the medium boundary $\Upsilon$ breaks down
the symmetry of random walks in its vicinity, which is reflected in the
anomalous behavior of the means quantities~\eqref{sec0:6a}--\eqref{sec0:6c}
near the boundary $\Upsilon$. To quantify this effect it is necessary to
calculate the given integrals near the boundary $\Upsilon$ for any small time
interval $\tau$.

Quantities~\eqref{sec0:6a}--\eqref{sec0:6c} comprise two type terms differing
in scaling with respect to $\tau$; regular components proportional to $\tau$
and anomalous one scaling as $\sqrt{\tau}$. In the present section only the
latter terms are under consideration. In deriving the Fokker-Planck equations
the division of them by $\tau$ gives rise to the singularity $(\tau)^{-1/2}$.
Exactly their cofactors quantify the influence of the boundary on Markovian
processes and setting them equal to zero we can relate the boundary values of
the Green function $G(\mathbf{r},t|\mathbf{r}_0,t_0)$ to the physical
properties of the medium boundaries.

Assuming the time scale $\tau$ to be sufficiently small the medium in a certain
neighborhood $\mathbb{Q}_\mathbf{s}$  of a boundary point
$\mathbf{s}\in\Upsilon$ is treated as a homogeneous continuum with time
independent characteristics and the corresponding fragment of the boundary
$\Upsilon$ is approximated by a hyperplane. In this case it is naturel to chose
the coordinate system related to a basis $\mathfrak{e}=
\mathfrak{e}_\Upsilon\oplus\boldsymbol{n}$, which, in particular, reduces the
number of the Green function arguments,
\[
G(\mathbf{r},x^M_0|\tau) :=
G\big(\mathbf{r},t_0+\tau|\{\mathbf{0}_\Upsilon,x^M_0\},t_0\big)\,.
\]
The system origin was located at the hyperplane $\Upsilon$ such that the vector
$\mathbf{r}_0 = \{\mathbf{0}_\Upsilon,x^M_0\}$ can have only one component
$x^M_0$ determining the distance between the point $\mathbf{r}_0$ and the
hyperplane $\Upsilon$. Then using the general
definitions~\eqref{sec0:6a}--\eqref{sec0:6c} of the quantities
$\mathfrak{R}(\mathbf{r},t,\tau)$, $\mathfrak{U}^i(\mathbf{r},t,\tau)$, and
$\mathfrak{L}^{ij}(\mathbf{r},t,\tau)$ the anomalous properties of random walks
near the boundary $\Upsilon$ are quantified by their singular components
$^*\mathfrak{U}^i(\tau,x^M)$ and $^*\mathfrak{L}^{ij}(\tau,x^M)$ scaling as
$\sqrt{\tau}$. The symbol $\ast$ is not applied to $\mathfrak{R}(\tau,x^M)$
because it possesses no regular component at all. In other words the desired
quantities are determined by the following means
\begin{align}
    \label{sing:1}
    &\int_{\mathbb{Q}_\mathbf{s}}
    d\tilde{\mathbf{r}}G(\tilde{\mathbf{r}},x^M|\tau)
    =1-
    \mathfrak{R}(\tau,x^M)\,,
\\
    \label{sing:2}
    &\int_{\mathbb{Q}_\mathbf{s}}d\tilde{\mathbf{r}}\,\delta \tilde{x}^i\,
    G(\tilde{\mathbf{r}},x^M|\tau)
    = {}^*\mathfrak{U}^i(\tau,x^M) +O(\tau)\,,
\\
    \label{sing:3}
    \frac12&\int_{\mathbb{Q}_\mathbf{s}}d\tilde{\mathbf{r}}\,
    \delta \tilde{x}^i \delta \tilde{x}^j\,
    G(\tilde{\mathbf{r}},x^M|\tau)
     = {}^*\mathfrak{L}^{ij}(\tau,x^M) +O(\tau)\,,
\end{align}
where $\delta \tilde{x}^\alpha = \tilde{x}^\alpha$ and $\delta \tilde{x}^M =
\tilde{x}^M-x^M$.

In order to calculate these boundary singularities we, first, fix the value
$\tau$ and introduce a new time scale $\tau_a\ll \tau$. Then the lattice
constructed is Sec.~\ref{sec2} and random walks on it are applied to calculate
the desired quantities. The advantage of using these lattice random walks is
due to two reasons. First, the choice of the basis $\mathfrak{b} =
\mathfrak{b}_\Upsilon\oplus\boldsymbol{b}_M$ enables us to simulate the
continuous Markovian process as independent random walks along the directions
parallel to the hyperplane $\Upsilon$ and along the vector $\boldsymbol{b}_M$.
Second, it becomes possible to ascribe special features to the nodes of the
boundary layer and in this way to simulate some physical properties of the
medium boundary. In particular, it either can absorb a random walker or causes
it to migrate extremely fast along the boundary within a thin layer. Finally,
to restor the continuous description the limit $\tau_a\to0$ is used.

The implementation of this approach again is based on just mathematical
manipulations with the probability function for lattice random walks. So only
the final results are stated here, referring a reader to
Appendix~\ref{appendix2} for the proof.

\begin{proposition}\label{app2.Prep3}

Let us consider a Markovian system in a homogeneous half-space
$\mathbb{R}^{M+}$ bounded by a hyperplane $\Upsilon$ and endowed with the basis
$\mathfrak{b}=\mathfrak{b}_\Upsilon\oplus\boldsymbol{b}_M$ described in
Proposition~\ref{app2.Prep2},
\[
    \mathbf{r} = \sum_{\gamma=1}^{M-1}\mathbf{b}_\gamma\zeta^\gamma + \boldsymbol{b}_M\zeta^M\,.
\]
The hyperplane $\Upsilon$ treated as a physical boundary can absorb the system
as well as force it to migrate fast along $\Upsilon$. The diffusion tensor
$D^{ij}$ as well as the drift velocity $v^i$ at the internal points and
$v^i_\Upsilon$ at the boundary $\Upsilon$ are assumed to be determined in the
basis $\mathfrak{b}$. It should be noted that the boundary drift velocity
$v^i_\Upsilon$ is the velocity at which the system had moved outside the
boundary if it would affected by the same forces.

The continuous motion of the Markovian system is imitated by random walks on
the lattice constructed in Sec.~\ref{sec2} with time step $\tau_a$. Finally the
limit $\tau_a\to0$ is applied.

Then, first, the boundary absorption and fast transport can be characterized by
two kinetic coefficients called the surface absorption rate $\sigma$ and the
surface diffusion length $l_\Upsilon$ ascribed directly to the boundary
$\Upsilon$ itself, meaning these quantities to be independent of the
discretization time $\tau_a$.

Second, random walks near the hyperplane $\Upsilon$ exhibit anomalous
properties reflected in the following singular means scaling with the time
$\tau$ as $\sqrt{\tau}$:
\begin{align}
    \mathfrak{R}_b(\tau,\zeta^M) &=D_{MM}^{-1/2}\sigma\cdot\mathcal{K}(\tau,\zeta^M)\,,
    \label{sing:4}
\\
    ^*\mathfrak{U}_{b}^{M}(\tau,\zeta^M) &= D_{MM}^{-1/2}\omega
    \cdot\mathcal{K}(\tau,\zeta^M)\,,
    \label{sing:5}
\\
    ^*\mathfrak{U}_{b}^{\alpha }(\tau,\zeta^M) &=
    D_{MM}^{-1/2}l_\Upsilon v_\Upsilon^{\alpha}\cdot
    \mathcal{K} (\tau,\zeta^M)\,,
    \label{sing:6}
\\
    ^*\mathfrak{L}_{b}^{\alpha \beta }(\tau,\zeta^M) &=
    D_{MM}^{-1/2}l_\Upsilon \mathcal{D}_\alpha \delta _{\alpha \beta }
    \cdot\mathcal{K}(\tau,\zeta^M)
    \,.
    \label{sing:7}
\end{align}
Here the label $b$ notes the basis $\mathfrak{b}$ used, $\zeta^M$ is the
distance between the point $\mathbf{r}$ and the hyperplane $\Upsilon$ measured
along the vector $\boldsymbol{b}_M$, and the function
$\mathcal{K}(\tau,\zeta^M)$ is specified by the integral
\begin{equation}\label{sing:8}
    \mathcal{K}(\tau,\zeta^M) = \sqrt{\frac{\tau}{\pi}} \int\limits_0^1
    \frac{dz}{\sqrt{z}}
    \exp\left[-\frac{(\zeta^M)^2}{4\mathcal{D}_M \tau}\,\frac1z\right]\,.
\end{equation}
\end{proposition}
In order to represent these boundary singularities in the initial basis
$\mathfrak{e}$ Proposition~\ref{app2.Prep2} is applied again. The initial basis
has been assumed to be of the form
$\mathfrak{e}=\mathfrak{e}_\Upsilon\oplus\boldsymbol{n}$ with the unit normal
$\boldsymbol{n}$ to the boundary $\Upsilon$ directed inwards the medium. Let
$\hat{U}_\Upsilon^{-1}=\|\breve{u}^\alpha_{\ \beta}\|$ be operator mapping the
boundary basis $\mathfrak{e}_\Upsilon$ onto the basis $\mathfrak{b}_\Upsilon$.
Then transition from the coordinates $\{\zeta^\alpha\},\zeta^M$ of a vector
$\mathbf{r}$ in the basis $\mathfrak{b}$ to its coordinates $\{x^\alpha\},x^M$
in the basis $\mathfrak{e}$ is specified by expressions~\eqref{app2.Prep2.D5b}
and \eqref{app2.Prep2.D5bM} using the tensor $\breve{u}^\alpha_{\ \beta}$ and
the diffusion tensor $D^{ij}$ determined in the \emph{initial} basis
$\mathfrak{e}$. In the vector form these coordinates are related by
equality~\eqref{add.13}. The quantities $^*\mathfrak{U}_{b}^i(\tau,\zeta^M)$
and $^*\mathfrak{L}_{b}^{ij}(\tau,\zeta^M)$ are obtained by averaging
variations of the coordinates $\zeta^i$. Thereby, they are a contravariant
vector and tensor, respectively, with the latter being proportional to the
diffusion tensor written in the basis $\mathfrak{b}$ and reduced to the
hyperplane $\Upsilon$, namely, the tensor $\mathcal{D}_a\delta_{\alpha\beta}$.
The value $\mathfrak{R}_b(\tau,\zeta^M)$ is naturally a scalar.  Whence it
follows directly that
\begin{align}
    \mathfrak{R}(\tau,x^M) &=D^{-1/2}_{MM}\sigma\cdot\mathcal{K}(\tau,x^M)\,,
    \label{sing:9}
\\
    ^*\mathfrak{U}^{i}(\tau,x^M) &= D^{-1/2}_{MM}\Big[
    D^{iM} + l_\Upsilon v_\Upsilon^i\Big]\cdot\mathcal{K}(\tau,x^M)\,,
    \label{sing:10}
\\
    ^*\mathfrak{L}^{\alpha \beta }(\tau,x^M) &=
    D^{-1/2}_{MM} l_\Upsilon \mathfrak{D}^{\alpha \beta}
    \cdot\mathcal{K}(\tau,x^M)
    \,.
    \label{sing:11}
\end{align}
Here the coordinate $x^M$ and $\zeta^M$ are interrelated by
formula~\eqref{app2.Prep2.D5bM} and the boundary diffusion tensor
$\mathfrak{D}^{\alpha\beta}$ is specified by expression~\eqref{IamG1}.
Formula~\eqref{sing:10} can be also rewritten in the vector form
\begin{equation}\label{sing:13}
        ^*\boldsymbol{\mathfrak{U}}(\tau,x^M) =D^{-1/2}_{MM}\Big[
        \boldsymbol{b} + l_\Upsilon \mathbf{v}_\Upsilon\Big]\mathcal{K}(\tau,x^M)\,,
\end{equation}
where the vector $\boldsymbol{b}$ of boundary singularities is given by
formula~\eqref{sec0:bsv1}.

\section{Boundary singularities and the boundary conditions}\label{sec4}

The obtained expressions~\eqref{sing:9}--\eqref{sing:11} actually directly lead
us to the final results. First, they relate the singular kinetic coefficients
to the diffusion tensor and the physical characteristics of the medium
boundary. Second, they reduce the problem of canceling the singularities inside
a think layer $\Upsilon_\tau$ adjacent the boundary $\Upsilon$ which,
nevertheless, is volumetric before implementing the passage to the limit
$\tau\to0$. Indeed, since all of terms~\eqref{sing:9}--\eqref{sing:11} depend
on the coordinate $x^M$ in the normal direction via the same function
$\mathcal{K}(\tau,x^M)$ the singularities will be canceled at all the points of
the layer $\Upsilon_\tau$ if it is the case at the boundary $\Upsilon$.
Besides, the structure of the function $\mathcal{K}(\tau,x^M)$, namely,
expression~\eqref{sing:8} justifies the adopted before assumption that the
characteristic thickness of the layer $\Upsilon_\tau$ scales with time as
$\tau^{1/2}$.

\subsection{Boundary condition for the backward Fokker-Planck equation}

As shown in Sec.~\ref{sec0:sub} the boundary singularities that appear in the
expansion the Chapman-Kolmogorov equation leading to the backward Fokker-Planck
equation will vanish if equality~\eqref{sec0:5BK} holds. At first, in order to
lighten the perception of results let us consider a rather small neighborhood
of the point $\mathbf{s}$ belonging to the boundary $\Upsilon$ wherein it is
actually a hyperplane and chose the basis
$\mathfrak{e}_\Upsilon\oplus\,\boldsymbol{n}$  composed of its hyperplane basis
$\mathfrak{e}_\Upsilon(\mathbf{s})$ and unit normal
$\boldsymbol{n}(\mathbf{s})$ directed inwards the domain $\mathbb{Q}$. Then
substituting expressions~\eqref{sing:9}--\eqref{sing:11} into
formula~\eqref{sec0:5BK} we immediately get the conclusion that at the boundary
point $\mathbf{s}\in\Upsilon$ the Green function
$G(\mathbf{r},t|\mathbf{r}_0,t_0)$ with respect to the latter pair of its
arguments with $\mathbf{r}_0\to\mathbf{s}$ has to meet the condition
{\setlength{\multlinegap}{0pt}
\begin{multline}\label{sec4:1}
    \sum_{i=1}^M D^{iM}\!(\mathbf{s},t_0)\nabla_i^s G(\mathbf{r},t|\mathbf{s},t_0)
    =
    \sigma(\mathbf{s},t_0)\, G(\mathbf{r},t|\mathbf{s},t_0)
\\
\begin{split}
    {} - l_\Upsilon(\mathbf{s},t_0)\bigg[&\sum_{\alpha = 1}^{M-1}
    v_\Upsilon^\alpha(\mathbf{s},t_0)
    \nabla_\alpha^s \,G(\mathbf{r},t|\mathbf{s},t_0)
\\
    {}+&\sum_{\alpha,\beta = 1}^{M-1}
    \mathfrak{D}^{\alpha\beta}(\mathbf{s},t_0)\nabla_\alpha^s\nabla_\beta^s
    \,G(\mathbf{r},t|\mathbf{s},t_0)\bigg].
\end{split}
\end{multline}
}%
We note that two last terms in expression~\eqref{sec4:1} describe effective
motion of the system inside the boundary $\Upsilon$ and has the form of the
backward Fokker-Planck operator~\eqref{BFPE2} with the diffusion tensor
$\mathfrak{D}^{\alpha\beta}$ and drift velocity $v_\Upsilon^\alpha$ whose
action is confined to the boundary $\Upsilon$. In order to rewrite this
expression for a orthonormal basis of general orientation we make use of the
definition of the boundary singularity vector $\boldsymbol{b}(\mathbf{s},t)$,
expression~\eqref{sec0:bsv1}, and take into account expression~\eqref{IamG1}
for the surface tensor diffusion. Then introducing the backward Fokker-Planck
operator acting only within the hyperplane $\Upsilon$
\begin{multline}\label{sec4:2}
    \bfp(\mathbf{s},t_0)\big\{\lozenge\big\} = l_\Upsilon(\mathbf{s},t_0)
\\
    {}\times
    \bigg[\sum_{i,j = 1}^M
    \mathfrak{D}^{ij}(\mathbf{s},t_0)\nabla_i^s\nabla_j^s
    \,\lozenge +
    \sum_{i = 1}^M  v_\Upsilon^i(\mathbf{s},t_0)
    \nabla_i^s \,\lozenge\bigg],
\end{multline}
where as before the symbol $\lozenge$ stands for a function acted by this
operator. Then in the vector invariant form the boundary condition for the
backward Fokker-Planck equation is written as
\begin{multline}\label{sec4:1g}
    \boldsymbol{b}(\mathbf{s},t_0)\cdot\boldsymbol{\nabla}^s
    G(\mathbf{r},t|\mathbf{s},t_0)
    =
    \sigma(\mathbf{s},t_0)\, G(\mathbf{r},t|\mathbf{s},t_0)
\\
    {} - \bfp(\mathbf{s},t_0)
    \big\{G(\mathbf{r},t|\mathbf{s},t_0)\big\}
\end{multline}
which is desired formula.

In deriving expression~\eqref{sec4:1} the boundary $\Upsilon$ was treated as a
hyperplane, i.e. the Euclidian space of dimension $(M-1)$ and its local basis
$\mathfrak{e}_\Upsilon$ was used. To write it again in the general form
underlining the fact that the operator $\bfp$ acts in this hyperplane only the
tensor notions of covariant derivatives are used (see, e.g.,
Ref.~\cite{tensor1}). In these terms the action of the operator $\bfp$ on the
Green function taken at the boundary $\Upsilon$ can be rewritten as
\begin{multline}\label{sec4:2a}
    \bfp(\mathbf{s},t_0)\big\{G(\mathbf{r},t|\mathbf{s},t_0)\big\}
     = l_\Upsilon(\mathbf{s},t_0)
\\
    {}\times
      \bigg[\sum_{\alpha=1}^{M-1}
    v_\Upsilon^\alpha(\mathbf{s},t_0)
    G(\mathbf{r},t|\mathbf{s},t_0)_{;\alpha}
    \\{}
    + \sum_{\alpha,\beta=1}^{M-1}
    \mathfrak{D}^{\alpha\beta}(\mathbf{s},t_0)
    \,G(\mathbf{r},t|\mathbf{s},t_0)_{;\alpha\beta}\bigg].
\end{multline}
In the given case it is no more that another form of the corresponding term in
expression~\eqref{sec4:1}. However, for a nonplanar boundary
formula~\eqref{sec4:2a} holds allowing for the boundary curvature, where as
expression~\eqref{sec4:2} loses the curvature effect. Its analysis goes far
beyond the scope of the present paper, so, here we will just ignore it.

\subsection{Boundary condition for the forward Fokker-Planck equation}

The boundary conditions are obtained in a similar way. First, we note that the
integrand of expression~\eqref{sec00:10} is similar to the boundary
relation~\eqref{sec0:5BK} within the replacement the test function
$\phi(\mathbf{r})$ by the Green function $G(\mathbf{r},t|\mathbf{r}_0,t_0)$ and
action of the operators at the argument $\mathbf{r}$ in stead of
$\mathbf{r}_0$. This analogy and the boundary condition~\eqref{sec4:1g} for the
backward Fokker-Planck equation enable us to reduce equality~\eqref{sec00:10}
to the following
\begin{equation}\label{sec40:1g}
    \boldsymbol{b}(\mathbf{s},t)\cdot\boldsymbol{\nabla}^s
    \phi(\mathbf{s})
    =
    \sigma(\mathbf{s},t)\phi(\mathbf{s})
     - \bfp(\mathbf{s},t)
    \big\{\phi(\mathbf{s})\big\}
\end{equation}
for an arbitrary boundary point $\mathbf{s}\in \Upsilon$. Since the boundary
part of the backward Fokker-Planck equation acts only within the boundary
$\Upsilon$ only the left part of expression~\eqref{sec40:1g} contains the first
derivative of the test function $\phi(\mathbf{s})$ in the direction normal to
the boundary $\Upsilon$ at the point $\mathbf{s}$. All the other terms are
either the boundary value of the function $\phi(\mathbf{s})$ itself or its
derivatives along the hyperplane $\Upsilon$. It justifies the adopted
previously statement that in the vicinity of $\Upsilon$ the test function
$\phi(\mathbf{r})$ can have any boundary value $\phi(\mathbf{s})$.

Then noting that the left-hand side of the condition~\eqref{sec00:2aa} is just
the combination
\[
    \boldsymbol{b}(\mathbf{s},t)\cdot\boldsymbol{\nabla}^s
    \,G(\mathbf{s},t|\mathbf{r}_0,t_0)
\]
the last equability converts expression~\eqref{sec00:2aa} into
\begin{multline}\label{sec40:2aa}
    \oint\limits_{\Upsilon} d\mathbf{s}
    \phi(\mathbf{s})    \sum_{i=1}^{M} \nu^{i}(\mathbf{s})
    \J^i\big\{G(\mathbf{s},t|\mathbf{r}_{0},t_{0})\}\,,
\\
\begin{split}
    {}= & - \oint\limits_{\Upsilon} d\mathbf{s} \phi(\mathbf{s})
    \sigma(\mathbf{s},t)\,G(\mathbf{s},t|\mathbf{r}_0,t_0)
\\
     {}&+\oint\limits_{\Upsilon} d\mathbf{s}
     \bfp(\mathbf{s},t)
    \big\{\phi(\mathbf{s})\big\} \,G(\mathbf{s},t|\mathbf{r}_0,t_0)\,.
\end{split}
\end{multline}
The last term in \eqref{sec40:2aa} using the divergence integral theorem for
the surfaces is reduced to the form
\begin{multline*}
      \oint\limits_{\Upsilon} d\mathbf{s}
    \bfp(\mathbf{s},t)
    \big\{\phi(\mathbf{s})\big\} \,G(\mathbf{s},t|\mathbf{r}_0,t_0)
    \\
    {} = \oint\limits_{\Upsilon} d\mathbf{s}
    \phi(\mathbf{s})
    \ffp(\mathbf{s},t)\big\{ G(\mathbf{s},t|\mathbf{r}_0,t_0) \big\}
    \,.
\end{multline*}
Here the operator $\ffp$ is the boundary forward Fokker-Planck equation
\begin{multline}\label{sec40:2}
    \ffp(\mathbf{s},t)\big\{\lozenge\big\}
     =\sum_{i= 1}^M \nabla_i^s
\\
     \times\bigg[\sum_{j = 1}^M
    \nabla_j^s\bigg(l_\Upsilon(\mathbf{s},t)\mathfrak{D}^{ij}(\mathbf{s},t)
    \,\lozenge\bigg) -
     l_\Upsilon(\mathbf{s},t)v_\Upsilon^i(\mathbf{s},t_0)
    \,\lozenge\bigg]\,,
\end{multline}
where again the symbol $\lozenge$ stands for the acted function. Since the test
function $\phi(\mathbf{s})$ takes any arbitrary values at the boundary
$\Upsilon$ equality~\eqref{sec40:2aa} holds for any point on the boundary
$\Upsilon$, i.e.
\begin{multline}\label{sec40:24}
    \boldsymbol{n}(\mathbf{s})\cdot
    \JJ\big\{G(\mathbf{s},t|\mathbf{r}_{0},t_{0})\} =
    -\sigma(\mathbf{s},t)\,G(\mathbf{s},t|\mathbf{r}_0,t_0)
\\
     {}+
     \ffp(\mathbf{s},t)
    \big\{G(\mathbf{s},t|\mathbf{r}_0,t_0)\big\}
\end{multline}
which is the desired boundary condition for the forwards Fokker-Planck
equation. As it should be the boundary condition~\eqref{sec40:24} can be
interpreted in terms of mass conservation; the component of walker flux normal
to the boundary $\Upsilon$ is determined by the surface rate of walker
absorption and the rate of fast surface transport withdrawing the walkers from
the given boundary point.

\section{Conclusion}

\begin{figure}
\begin{center}
\includegraphics[width=0.9\columnwidth]{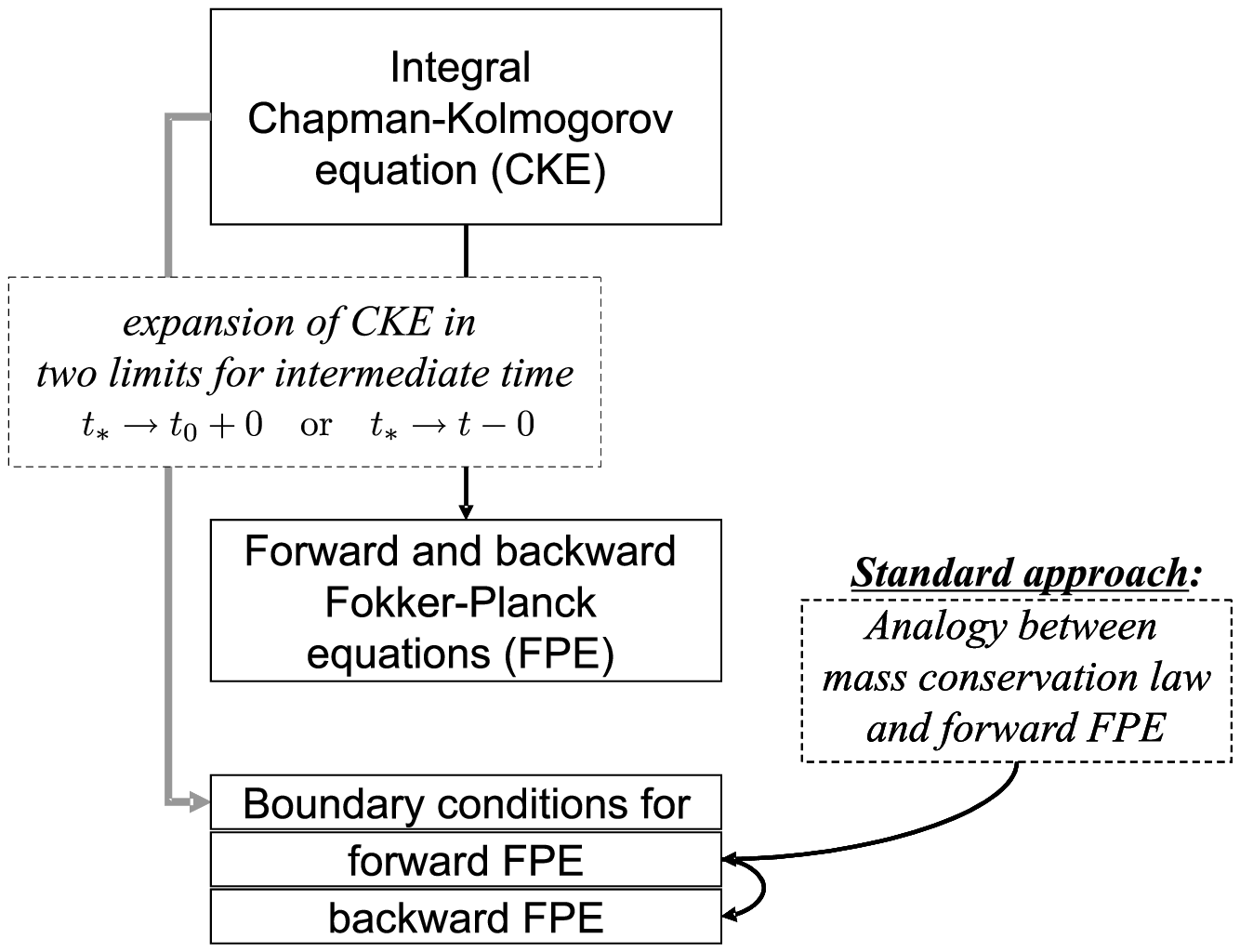}
\caption{Illustration of the main purpose of the present work represented by
grey directed line.} \label{LFig}
\end{center}
\end{figure}

The present paper has developed a technique of \textit{deriving} the boundary
conditions for the Fokker-Planck equations based on the Chapman-Kolmogorov
integral equation. The idea of the work is illustrated in Fig.~\ref{LFig}.

The interest to this problem is partly due to the following. It is well known
that the Fokker-Planck equations, forward and backward ones, stem directly from
the Chapman-Kolmogorov equation under additional two assumptions, the short
time confinement of the corresponding Markovian process and the local
homogeneity of the medium. There are rather rigorous techniques of deriving
them from the integral Chapman-Kolmogorov equation based on expanding the
latter on short time scales in the possible limits. By contrast, the
corresponding boundary conditions are typically postulated applying to the
physical meaning of the probability flux and the analogy between the forward
Fokker-Planck equation and the mass conservation law.

However such simple arguments can fail in dealing with more complex Markovian
processes like sub- or super-diffusion, for which the Fokker-Planck equations
with fractional derivatives form the governing equations. In this case it would
be appropriate to have a formal technique giving rise to the boundary
conditions starting from the general description. However, up to now
constructing such a technique is a challenging problem. It was the case also
with respect to the normal Markovian processes in  continua.

This paper actually has demonstrated how to do this dealing with the normal
Markovian processes. The key point is the fact that the medium boundary breaks
down the symmetry of random walks near it. As result, the coefficients in the
corresponding expansion series of the Chapman-Kolmogorov equation are endowed
with anomalous features called the boundary singularities. Namely, they scale
on short time scales as $\delta t^{-1/2}$. Since the probability distribution
on macroscopic scales cannot contain such singularities the corresponding
cofactors in the expressions for the boundary singularities should be set equal
to zero, leading one to the required boundary conditions. In this way we have
shown that the boundary conditions of the Fokker-Planck equations are also the
direct consequence of the Chapman-Kolmogorov equation supplemented with some
rather general assumptions about the properties of the medium boundary. As it
must the boundary conditions obtained in this way match mass conservation.

%

\appendix

\section{Proof of Proposition~\ref{app2.Prep2}}\label{appendix}

At the first step the initial basis $\mathfrak{e}$ of the half-space
$\mathbb{R}^{M+}$ is assumed to comprise a certain basis
$$
\mathfrak{e}_{\Upsilon}=\{\mathbf{e}_1,\mathbf{e}_2,\ldots\mathbf{e}_{M-1}\}
$$
of the hyperplane $\Upsilon$ and its unit normal $\boldsymbol{n}$ directed
inward $\mathbb{R}^{M+}$, i.e. $\mathfrak{e} = \mathfrak{e}_\Upsilon \oplus
\boldsymbol{n}$. Then the results to be obtained will be represented in
invariant form where appropriate, enabling us to write the general expressions.
The diffusion tensor $D^{ij}$ is regarded to be determined beforehand in the
initial basis. Besides, as before, to simplify reader perception the Greek
letters will be used to label tensor indices corresponding to the hyperplane
$\Upsilon$.

Let us consider a new basis $\mathfrak{b} = \mathfrak{b}_\Upsilon \oplus
\boldsymbol{b}_M$ of the same structure except for the last vector
$\boldsymbol{b}_M$; it can be not normal to the hyperplane $\Upsilon$. An
one-to-one map between the two bases,
$\mathfrak{e}\Leftrightarrow\mathfrak{b}$, determines a linear transformation
$\widehat{\mathcal{U}}$ of the space $\mathbb{R}^M$ mapping, in particular, the
hyperplane $\Upsilon$ onto itself. This transformation
$\widehat{\mathcal{U}}=\|U^i_{\ j}\|$ is specified by the relationship between
the basis vectors
\begin{equation}\label{app2.20}
 \mathbf{e}_{\alpha} = \sum_{\beta=1}^{M-1}\mathbf{b}_{\beta} u^\beta_{\ \alpha}
 \quad\text{è}\quad
 \boldsymbol{n} = \sum_{\alpha=1}^{M-1}\mathbf{b}_{\alpha}\omega^\alpha + \boldsymbol{b}_M\omega^M.
\end{equation}
Here the tensor $u^\alpha_{\ \beta}$ represents an operator
$\widehat{\mathcal{U}}_\Upsilon$ acting in the hyperplane $\Upsilon$ whereas
the tensor $\omega^\alpha$ (in $\Upsilon$) and the coefficient $\omega^M\neq 0$
complement it to the operator $\widehat{\mathcal{U}}$, namely,
\begin{subequations}\label{app2.21}
\begin{align}
    \label{app2.21b}
    U^\alpha_{\ \beta} & = u^\alpha_{\ \beta}\,,
    & U^\alpha_{\ M}& =
    \omega^\alpha\,,
    \\
    \label{app2.21a}
    U^M_{\phantom{M} \beta} & = 0\,, &U^M_{\phantom{M} M} & = \omega^M.
\end{align}
\end{subequations}
According to the rule of tensor transformations (see, e.g.,
Ref.~\cite{tensor1}) in the basis $\mathfrak{b}$ the diffusion matrix has the
components
\begin{subequations}\label{app2.22}
\begin{align}
    \label{app2.22c}
    \begin{split}
    \Tilde{D}^{\alpha\beta} & = \sum_{\gamma,\gamma'=1}^{M-1}
    u^\alpha_{\ \gamma} u^\beta_{\ \gamma'}D^{\gamma\gamma'}
    +
    \omega^\alpha \omega^\beta D^{MM}
    \\{}&\quad{}+ \sum_{\gamma=1}^{M-1}
    \left(\omega^\alpha u^\beta_{\ \gamma}+
    \omega^\beta u^\alpha_{\ \gamma}\right) D^{\gamma M}\,,
    \end{split}
    \\
    \label{app2.22b}
    \Tilde{D}^{\alpha M} & = \omega^M \bigg(
    \sum_{\gamma=1}^{M-1}u^\alpha_{\ \gamma}D^{\gamma M}
    +\omega^\alpha D^{MM}\bigg)\,,
    \\
    \label{app2.22a}
    \Tilde{D}^{MM} & = \left(\omega^M\right)^2 D^{MM} \,.
\end{align}
\end{subequations}
Correspondingly, an arbitrary vector $x^i=\{x^\alpha, x^M\}$ is converted as
\begin{subequations}\label{app2.22v}
\begin{align}
    \label{app2.22cv}
    \Tilde{x}^{\alpha} & = \sum_{\gamma=1}^{M-1}
    u^\alpha_{\ \gamma} x^{\gamma}
    +
    \omega^\alpha x^{M}\,,
    \\
    \label{app2.22av}
    \Tilde{x}^{M} & = \omega^M x^{M} \,.
\end{align}
\end{subequations}
Currently there is no restrictions imposed on the basis $\mathfrak{b}$ (except
for its general structure). Now let us choose a specific version of the tensor
$\omega^\alpha$ that eliminates the off-diagonal elements of the diffusion
tensor in the basis $\mathfrak{b}$. By virtue of \eqref{app2.22b} it is
\begin{equation}\label{app2.23}
    \omega^\alpha = -\frac1{D^{MM}}\sum_{\gamma=1}^{M-1}u^\alpha_{\ \gamma}D^{\gamma M}\,.
\end{equation}
Here the division by $D^{MM}$ is possible because according to
definition~\eqref{app2.1} the diagonal elements of diffusion tensor are
positive, in particular, $D^{MM}> 0$ except for the case where the system
motion along the direction $\boldsymbol{n}$ is rigorously deterministic.
However, setting further $D^{MM}\to+0$ the latter case can be also allowed for.
The substitution of \eqref{app2.23} into \eqref{app2.22c} yields
\begin{equation}
  \label{app2.24}
    \Tilde{D}^{\alpha\beta}  = \sum_{\gamma,\gamma'=1}^{M-1}
    u^\alpha_{\ \gamma}u^\beta_{\ \gamma'} \mathfrak{D}^{\gamma\gamma'}\,,
\end{equation}
where the object
\begin{equation}\label{app2.25}
  \mathfrak{D}^{\alpha\beta}  = D^{\alpha\beta} - \frac1{D^{MM}}\,D^{\alpha M}D^{\beta
  M}
\end{equation}
is a tensor within the hyperplane $\Upsilon$ because, up to now, the
collections of vectors $\mathfrak{e}_\Upsilon$ and $\mathfrak{b}_\Upsilon$ are
general bases of this hyperplane.

The tensor $\mathfrak{D}^{\alpha\beta}$ is symmetric and positive definite. The
latter property stems directly from inequality~\eqref{app.new2} written for an
arbitrary covector $l_\alpha$ of the hyperplane $\Upsilon$ with the component
\begin{gather}
  l_M = -\frac{1}{D^{MM}}\sum_{\gamma=1}^{M-1}D^{M\gamma}l_{\gamma}\,,\\
\intertext{namely,}
  \sum_{i,j=1}^{M}D^{ij}l_il_j = \sum_{\alpha,\beta=1}^{M-1}
  \mathfrak{D}^{\alpha\beta}l_\alpha l_\beta > 0\,.
\end{gather}
Therefore the basis $\mathfrak{b}_\Upsilon$ of the hyperplane $\Upsilon$ can be
chosen to be orthonormal one wherein the tensor $\mathfrak{D}^{\alpha\beta} $
takes the diagonal form with the diagonal components being positive values,
i.e. $\mathfrak{D}^{\alpha\beta} = \mathfrak{D}_\alpha^\beta =
\mathfrak{D}_{\alpha\beta} = \mathcal{D}_{\alpha}\delta_{\alpha\beta}$
\cite{Gantmaher}. This basis $\mathfrak{b}_\Upsilon$ is made up of the
eigenvectors of the operator
$\widehat{\mathfrak{D}}:=\|\mathfrak{D}^\alpha_\beta\|$ whose eigenvalues are
$\{D_\alpha\}$. For example, in the initial basis $\mathfrak{e}_\Upsilon$ the
tensor $\mathfrak{D}_\alpha^\beta$ is related to the tensor
$\mathfrak{D}^{\alpha\beta}$ by the expression
\begin{equation*}
   \mathfrak{D}_\alpha^\beta = \sum_{\gamma=1}^{M-1} g_{\alpha\gamma}
   \mathfrak{D}^{\gamma\beta},
   \quad\text{where}\quad
   g_{\alpha\beta}:=(\mathbf{e}_\alpha\cdot\mathbf{e}_\beta)
\end{equation*}
is the matric tensor of the hyperplane $\Upsilon$.

The choice of the given basis $\mathfrak{b}_\Upsilon$ specifies the
transformation matrix $\|u^\alpha_{\ \beta}\|$ which together with
expression~\eqref{app2.23} gives us the vector $\boldsymbol{b}_M$ and the
corresponding component $\mathcal{D}_M$ of the diffusion tensor. Namely, first,
substituting \eqref{app2.23} into the latter equality of \eqref{app2.20} and
taking into account the former one we write
\begin{align}\nonumber
 \boldsymbol{b}_M\omega_M & = \boldsymbol{n} + \frac1{D^{MM}}\sum_{\alpha,\gamma=1}^{M-1}
 \mathbf{b}_\alpha u^\alpha_{\ \gamma} D^{\gamma M}\\
 \label{app.new3_1}
 {}& = \boldsymbol{n} + \frac1{D^{MM}}\sum_{\gamma=1}^{M-1}\mathbf{e}_\gamma D^{\gamma M}\,.
\end{align}
In the invariant form this expression can be rewritten as
\begin{gather}
    \label{app.new3}
    \boldsymbol{b}_M  = \frac1\omega \sum_{i,j=1}^{M}
    \mathbf{e}_iD^{ij}(\mathbf{e}_j\cdot\boldsymbol{n})\,,\\
\intertext{where the normalization factor $\omega$}
    \label{app.new4}
    \omega  = \bigg[\sum_{i,j,k,p=1}^{M}
    D^{ik}D^{jp}(\mathbf{e}_i\cdot\mathbf{e}_j)(\mathbf{e}_k\cdot\boldsymbol{n})
    (\mathbf{e}_p\cdot\boldsymbol{n})
    \bigg]^{1/2}
\end{gather}
is due to the vector $\boldsymbol{b}_M$ being of unit length. Since the
obtained expressions~\eqref{app.new3} and \eqref{app.new4} are of the tensor
form and are scaler in this meaning they hold within any basis, proving
formulae~\eqref{app2.Prep2.D1} and \eqref{app2.Prep2.D2}.

Second, according to \eqref{app.new3_1} and \eqref{app.new3} the coefficient
$\omega_M = \omega/D^{MM}$. Thereby expressions~\eqref{app2.22a} and
\eqref{app.new4} give us the diffusion tensor component $\mathcal{D}_M$ related
to the vector $\boldsymbol{b}_M$ in the basis $\mathfrak{b}$
\begin{equation}\label{app.new5}
    \mathcal{D}_M = \omega^2 \bigg[\sum_{i,j=1}^{M-1}
    D^{ij}(\mathbf{e}_i\cdot\boldsymbol{n})(\mathbf{e}_j\cdot\boldsymbol{n})
    \bigg]^{-1}
\end{equation}
and written in the invariant form. Formula~\eqref{app2.Prep2.D4} is proved. In
addition, expressions~\eqref{app2.22v} and \eqref{app2.23} together with the
equality $\omega = \omega_M D^{MM}$ immediately lead to
formulae~\eqref{app2.Prep2.D5a} and \eqref{app2.Prep2.D5aM}.

Finally, we need the transformation
$\widehat{\mathcal{U}}_\Upsilon^{-1}=\|\breve{u}^\alpha_{\ \beta}\|$ of the
hyperplane $\Upsilon$ that is inverse to the transformation
$\widehat{\mathcal{U}}_\Upsilon = \|u^\alpha_{\ \gamma}\|$; its components obey
the equality
\begin{equation}\label{app.new8}
    \sum_{\gamma=1}^{M-1} \breve{u}^\alpha_{\ \gamma}
    u^\gamma_{\ \beta} = \delta^\alpha_\beta\,.
\end{equation}
It exists due to the transformation $\widehat{\mathcal{U}}_\Upsilon$ being
one-to-one map of the bases $\mathfrak{e}$ and $\mathfrak{b}$. Then the
inversion of equalities~\eqref{app2.22v} with $\omega^\alpha$ given by
expression~\eqref{app2.23} yields formulae~\eqref{app2.Prep2.D5b} and
\eqref{app2.Prep2.D5bM}. Inverting now relationship~\eqref{app2.24} and taking
into account the tensor $\Tilde{D}^{\alpha\beta}$ to have the diagonal form
$\mathcal{D}_\alpha\delta_{\alpha\beta}$ in the orthonormal basis
$\mathfrak{b}$ we directly get
\begin{equation*}\label{app.new7}
    \sum_{\gamma = 1}^{M-1}
    \breve{u}^\alpha_{\ \gamma}
    \breve{u}^\beta_{\ \gamma}
    \mathcal{D}_\gamma = \mathfrak{D}^{\alpha\beta}\,.
\end{equation*}
which together with expressions~\eqref{app2.25} gives rise to
formula~\eqref{app2.Prep2.D6}. The Proposition is proved.

\section{Proof of Proposition~\ref{app2.Prep3}}\label{appendix2}

The homogeneous half-space $\mathbb{R}^{M+}$ bounded by the hyperplane
$\Upsilon $ is under consideration and a lattice described in Sec.~\ref{sec2}
is constructed. It is made up of the node layers $\{\Upsilon _{i}\}$ parallel
to the hyperplane $\Upsilon $ with the interplane spacing vector
$a_M\boldsymbol{b}_M$. The individual node arrangement of the  layers $\Upsilon
_{i}$ is determined by the vectors of the hyperplane basis
$\mathfrak{b}_{\Upsilon }$ with spacings $\{a_{\alpha }\}$. In other words, the
nodes of this lattice are the points
\[
\mathbf{r}_{\mathbf{n}}=\sum_{\alpha =1}^{M-1}n^{\alpha }\left( a_{\alpha }%
\mathbf{b}_{\alpha }\right) +na_M\boldsymbol{b}_M\,,
\]
where $\mathbf{n}$ is the collection of numbers
$\left\{\mathbf{n}_{\Upsilon},n\right\} = \{\left\{ n^{\alpha }\right\},n\}$
taking any integer value, $\left. n^{\alpha }\right\vert _{1}^{M-1}=0,\pm 1,\pm
2,\ldots $, except for the last one; it takes only nonnegative values $
\,n=0,1,2,\ldots $ In particular, the points $\{\mathbf{r}_{\mathbf{n}
}\}_{\Upsilon }$ with $n=0$ form the boundary layer $\Upsilon _{0}$.

The Markovian process in the half-space $\mathbb{R}^{M+}$ is simulated by
random walks on this lattice with hop probabilities given in Sec.~\ref{sec2}.
To find the desired boundary singularities we will analyze evolution of the
walker distribution over the given lattice, i.e. the dynamics of the
probability $\mathcal{P}_{t,\mathbf{n}}$ to find the walker at node $
\mathbf{n}$ after hop $t$. Here $t$ is the time measured in jump numbers, i.e.
in units of the hop duration $\tau _{a}$. At the initial time $t=0$ the walker
is assumed to be located at a certain internal node $\mathbf{n}_{0}$. Without
lost of generality all the components of the index $\mathbf{n}_{0}$ can be set
equal to zero except for the last one, i.e. $\mathbf{n}_{0}=\{0,0,\ldots
,0,n_{0}\}$.

\subsection{Moments of the walker distribution and the generation function}

Actually the main purpose of the present appendix is to find the zero-th,
first, and second order moments of the distribution function
$\mathcal{P}_{t,\mathbf{m}}$. The zero-th moment quantifies the trapping
effect, whereas the fist and second ones characterize the walker propagation in
space. Namely, the following quantities
\begin{align}
    \mathfrak{R}_a(t,n_{0}) &=1-\sum\limits_{n=0}^{\infty }\sum_{\mathbf{n}
    _{\Upsilon}}\mathcal{P}_{t,\{\mathbf{n}_{\Upsilon },n\}}\,,
    \label{app2.v1}
    \\
    \mathfrak{U}_a^{i}(t,n_{0}) &=\sum\limits_{n=0}^{\infty }\sum_{\mathbf{n}
    _{\Upsilon }}\left( n^{i}-n_{0}^{i}\right) \mathcal{P}_{t,\{\mathbf{n}
    _{\Upsilon },n\}}\,,
    \label{app2.v2}
                    \\
    \mathfrak{L}_a^{ij}(t,n_{0}) &=\frac12\sum\limits_{n=0}^{\infty }\sum_{\mathbf{n}
    _{\Upsilon }}\left( n^{i}-n_{0}^{i}\right) \left( n^{j}-n_{0}^{j}\right)
    \mathcal{P}_{t,\{\mathbf{n}_{\Upsilon },n\}}
    \label{app2.v3}
\end{align}
have to be calculated. Here the index $i$ is used as a general symbol for one
of the indices $\{\alpha \},M$. In order to do this the generation function and
its analogy written for the boundary nodes only
\begin{align}
    G(s,p,\mathbf{k}_{\Upsilon }) &=\sum\limits_{\substack{t=0\\n=0}}^{\infty}
    \sum_{\mathbf{n}_{\Upsilon }}
    e^{
    -st-p(n-n_{0})+\mathrm{i}(\mathbf{k}_{\Upsilon }\cdot \mathbf{n}_{\Upsilon })
    }
    \,\mathcal{P}_{t,\{\mathbf{n}_{\Upsilon },n\}},
    \label{gf}
       \\
        \nonumber
    g(s,\mathbf{k}_{\Upsilon }) &=\sum\limits_{t=0}^{\infty }\sum_{\mathbf{n}
    _{\Upsilon }}
    e^{ -st+\mathrm{i}(\mathbf{k}_{\Upsilon }\cdot
    \mathbf{n}_{\Upsilon })
    }
    \,\mathcal{P}_{t,\{\mathbf{n}_{\Upsilon},0\}}
    \\
    {} &=
    \lim_{p\rightarrow \infty }\left[ e^{-pn_{0}}G(s,p,\mathbf{k} _{\Upsilon
    })\right]
    \label{gfbound}
\end{align}
are introduced, where the complex arguments $s$, $p$ have the positive real
parts, $\mathrm{Re\,}s,\mathrm{Re\,}p\geq 0$. It should be noted that the traps
are not included into these sums. The discrete Laplace transforms of the
desired functions \eqref{app2.v1}--\eqref{app2.v3} are directly related to the
generation function. Indeed
\begin{align}
    \mathfrak{R}_{a}(s,n_{0}) &= \sum\limits_{t=0}^{\infty
    }e^{-st}\mathfrak{R}_a
    (t,n_{0})=\frac{1}{\left( 1-e^{-s}\right) }-G(s,0,\mathbf{0})\,,
    \label{app2.vl1}
    \\
    \mathfrak{U}_{a}^{i}(s,n_{0}) &=\sum\limits_{n=0}^{\infty }e^{-st}
    \mathfrak{U}_a^{i}(t,n_{0})=\nabla _{i}\left. G(s,p,\mathbf{k}_{\Upsilon})
    \right\vert _{p,\mathbf{k}_{\Upsilon }=0}\,,
    \label{app2.vl2}
    \\
    \nonumber
    \mathfrak{L}_{a}^{ij}(s,n_{0}) &=\sum\limits_{n=0}^{\infty }e^{-st}
    \mathfrak{L}_a^{ij}(t,n_{0})\\
    &\qquad\quad{}=\frac12\nabla _{i}\nabla _{j}\left. G(s,p,\mathbf{k}
    _{\Upsilon })\right\vert _{p,\mathbf{k}_{\Upsilon }=0},
    \label{app2.vl3}
\end{align}
where the operator $\nabla _{i}$ is $\nabla _{\alpha }=-\mathrm{i}\partial
_{k^{\alpha }}$ if the index $i=\alpha $ is one of the indices of the
hyperplane $\Upsilon $ and $\nabla \,_{M}=-\partial _{p}$ for the index $i=M$.

\subsection*{Master equation for lattice random walks and its general solution}

To find the generation function for the discrete random walks under
consideration the corresponding master equation is applied. For an internal
node $\mathbf{n=}\left\{ \mathbf{n}_{\Upsilon },n\right\} $ with $n\geq 2$ it
takes the form
\begin{equation}
  \mathcal{P}_{t+1,\mathbf{n}}=\sideset{}{'}\sum_{\mathbf{m}}
  \mathcal{P}_{t,\mathbf{m}}\,P_{\mathbf{mn}}\,.
  \label{app20:eq.1}
\end{equation}
Here prime at the sum denotes the index $\mathbf{m}$ running over all the
nearest neighbors of the given node $\mathbf{n}$ and according to expression
\eqref{app2.33} the corresponding hop probabilities can be represented as
\begin{equation}
    P_{\mathbf{mn}}=\frac{1+\epsilon _{i}\chi _{i}}{2M}\,,
    \label{app20:eq.2}
\end{equation}
where $\epsilon _{i}=\tau _{a}v^{i}M/a_{i}$ are some small quantities scaling
with $\tau _{a}$ as $\epsilon_{i}\propto \tau _{a}^{1/2}$ and the value $\chi
_{i}=\pm 1$ stands for hops along the basis vector $\mathbf{b}_{i}$ or in the
opposite direction, i.e. the hop to the node with $m^{i}=n^{i}\pm 1 $ and
$m^{j}=n^{j}$ for $j\neq i$. For the nodes of the layer $\Upsilon _{1}$ the
master equation becomes
\begin{equation}
    \mathcal{P}_{t+1,\mathbf{n}}=\sideset{}{'}\sum_{\mathbf{m}}\mathcal{P}_{t,
    \mathbf{m}}\,P_{\mathbf{mn}}+\mathcal{P}_{t,\mathbf{n}_{b}}P_{l}\,.
    \label{app20:eq.3}
\end{equation}
Here again prime at the sum has the same meaning except for only internal
neighboring nodes being taken into account, $\{\mathbf{n}_{b},\mathbf{n}\}$ is
the pair of nodes belonging to the boundary layer $\Upsilon _{0}$ and the
adjacent internal layer $\Upsilon _{1}$ that are related to each other via
walker hops, and the hop probability $P_{l}$ is determined by expression
\eqref{app2.34a}. The walker distribution function $P_{\mathbf{n}_{b},t}$ in
the boundary layer obeys the equation
\begin{equation}
    \mathcal{P}_{t+1,\mathbf{n}_{b}}=\sum_{\mathbf{m}_{b}\in \Upsilon _{0}}
    \mathcal{P}_{t,\mathbf{m}_{b}}\,P_{\Upsilon }P_{\mathbf{m}_{b}\mathbf{n}
    _{b}}^{(g)}+\mathcal{P}_{t,\mathbf{n}}P_{\mathbf{nn}_{b}}\,.
    \label{app20:eq.4}
\end{equation}
We remind that the jumps inside the boundary layer can be complex and comprise
individually $g$ elementary hops. In this case the multihop probability
$P_{\mathbf{m}_{b}\mathbf{n}_{b}}^{(g)}$ is determined by formula
\eqref{app2.36}. The one-hope probability along the basis vector
$\mathbf{b}_{\alpha }$ provided the walker remains in the boundary layer
$\Upsilon _{0}$ is
\begin{equation}
    P_{\mathbf{m}_{b}\mathbf{n}_{b}}^{(1)}=
    \frac{1+\epsilon _{\alpha }^{\Upsilon}\chi _{\alpha }}{2(M-1)}\,,
    \label{add.app.1}
\end{equation}
where $\epsilon _{\alpha }^{\Upsilon }=\tau _{a}v_{\Upsilon
}^{\alpha}M/a_{\alpha }$ is again a small parameter scaling as  $\epsilon
_{\alpha }^{\Upsilon }\propto \tau _{a}^{1/2}$. The values $\epsilon_\alpha$
quantify the asymmetry of hops in the boundary layer $\Upsilon _{0}$. In
particular, these complex jumps are characterized by the means
\begin{align}
    \left\langle n^{\alpha }\right\rangle _{\Upsilon } & =\sum_{\mathbf{n}
    _{b}\in \Upsilon _{0}}n^{\alpha }P_{\mathbf{0n}_{b}}^{(g)}=\frac{g}{(M-1)}
    \,\epsilon _{\alpha }^{\Upsilon }\,,
\label{bcor1}
    \\
    \left\langle n^{\alpha }n^{\beta }\right\rangle _{\Upsilon } &=\sum_{
    \mathbf{n}_{b}\in \Upsilon _{0}}n^{\alpha }n^{\beta }
    P_{\mathbf{0n}_{b}}^{(g)}
\nonumber   \\
    {}&=\frac{g}{(M-1)}\,\delta _{\alpha \beta }+
    \frac{g(g-1)}{(M-1)^{2}}\,\epsilon _{\alpha }^{\Upsilon }\epsilon _{\beta }^{\Upsilon}
    \,. \label{bcor2}
\end{align}
Finally, the master equation for the traps is
\begin{equation}
    \mathcal{P}_{t+1,\mathbf{n}_{b}}^{(tr)}=
    \mathcal{P}_{t,\mathbf{n}_{b}}^{(tr)}
    +
    \mathcal{P}_{t,\mathbf{n}_{b}}\,P_{tr}\,.
    \label{app20:eq.6}
\end{equation}
The hop probabilities $P_{l}$, $P_{tr}$, are given by expressions
\eqref{app2.34a} and the kinetic coefficients of walker jumps inside the
boundary layer $\Upsilon _{0}$ are specified by expressions \eqref{app.new10},
\eqref{app2.35}, and \eqref{app2.36}. At the initial time the walker
distribution meets the condition
\begin{equation}
    \mathcal{P}_{t=0,\mathbf{n}}=\delta _{\mathbf{nn}_{0}}\,.
    \label{app20:eq.7}
\end{equation}

To solve this system of equations we substitute \eqref{app20:eq.1},
\eqref{app20:eq.3}, and \eqref{app20:eq.4} into definition \eqref{gf} of the
generation function $G(s,p,\mathbf{k}_{\Upsilon }) $ and after succeeding
mathematical manipulations get the following equation (see comments about its
derivation just after formula~\eqref{app20:eq.11})
\begin{multline}
    \left[ e^{s}-\Phi \left( p,\mathbf{k}_{\Upsilon }\right) \right] G\left(s,p,
    \mathbf{k}_{\Upsilon }\right)
    \\
    =e^{s}-e^{pn_{0}}
    \left[ \Phi \left( p,\mathbf{k}_{\Upsilon }\right) -\phi
    \left( p,\mathbf{k}_{\Upsilon }\right) \right] g(s,\mathbf{k}_{\Upsilon })
    \label{geq}
\end{multline}
relating the given generation functions $G(s,p,\mathbf{k}_{\Upsilon })$ and
$g(s,\mathbf{k}_{\Upsilon })$ to each other. Here the following functions
\begin{align}
    \Phi(p,\mathbf{k}_{\Upsilon }) & =
    \frac{1}{M}\left( \cosh p-\epsilon _{M}\sinh p\right)
\nonumber \\
    {}& + \frac{1}{M}\sum_{\alpha =1}^{M-1}\left( \cos k_{\alpha }+
    \mathrm{i}\epsilon _{\alpha }\sin k_{\alpha }\right)\,,
    \label{app20:eq.10}
\\
    \phi (p,\mathbf{k}_{\Upsilon }) &= \frac{\left( 1-\sigma _{a}\right) }{M} e^{-p}
\nonumber \\
    {}& +\frac{(M-1)}{M}\sum_{\alpha =1}^{M-1}\exp \left[ \mathrm{i}(\mathbf{k}
    _{\Upsilon }\cdot \mathbf{n}_{\Upsilon })\right] P_{\mathbf{0n}_{\Upsilon}}^{(g)}
    \label{app20:eq.11}
\end{align}
have been constructed in deriving equation~\eqref{geq}.
\vspace{0.5\baselineskip}

\noindent%
\textit{Comments on deriving equation \eqref{geq}} The key fragments of
deriving equation~\eqref{geq} are outlined below. The conversion in \eqref{gf}
from $t\rightarrow t+1$ leads to the line
\begin{gather*}
    G(s,p,\mathbf{k}_{\Upsilon }) =
    e^{-s}\mathcal{G}(s,p,\mathbf{k}_{\Upsilon })+1\,,
\\ \intertext{where}
    \mathcal{G}(s,p,\mathbf{k}_{\Upsilon })  =
    \sum\limits_{\substack{t=0\\n=0}}^{\infty}
    \sum_{\mathbf{n}_{\Upsilon }}
    e^{
    -st-p(n-n_{0})+\mathrm{i}(\mathbf{k}_{\Upsilon }\cdot \mathbf{n}_{\Upsilon })
    }
    \,\mathcal{P}_{t+1,\{\mathbf{n}_{\Upsilon },n\}}
\end{gather*}
and the initial condition \eqref{app20:eq.7} has been taken into account.
Equations~\eqref{app20:eq.1}, \eqref{app20:eq.3}, and \eqref{app20:eq.4}
relating two succeeding steps of random walks are substituted into the latter
expression. As a result the terms in
sums~\eqref{app20:eq.1}--\eqref{app20:eq.7} matching the interlayer hops split
it into two parts
\begin{multline*}
    \mathcal{G}(s,p,\mathbf{k}_{\Upsilon })\Rightarrow \Phi_1 (p)G(s,p,\mathbf{k}
    _{\Upsilon })
    \\
    {}+ e^{pn_{0}}\left[ \phi_1 (p)-\Phi_1 (p)\right] g(s,\mathbf{k}
    _{\Upsilon })
\end{multline*}
with the latter summand caused by that the boundary nodes differ from the
internal ones in properties. In their turn the components of
sums~\eqref{app20:eq.1}--\eqref{app20:eq.7} describing transitions between a
given node $\mathbf{n}$ and the nodes of the same layer also split the term
$\mathcal{G}(s,p,\mathbf{k}_{\Upsilon })$ into two parts
\begin{multline*}
    \mathcal{G}(s,p,\mathbf{k}_{\Upsilon })\Rightarrow
    \Phi_2 (\mathbf{k}_{\Upsilon })G(s,p,\mathbf{k}_{\Upsilon })\\
    {}+ e^{pn_{0}}
    \left[ \phi_2 (\mathbf{k}_{\Upsilon})-\Phi_2 (\mathbf{k}_{\Upsilon })\right]
    g(s,\mathbf{k}_{\Upsilon })\,,
\end{multline*}
where the latter summand is due to fast diffusion in the boundary layer. The
combination of the two last lines gives equation \eqref{geq} with $\Phi \left(
p,\mathbf{k}_{\Upsilon }\right) =\Phi_1 (p)+\Phi_2 (\mathbf{k}_{\Upsilon })$
and $\phi \left( p,\mathbf{k}_{\Upsilon }\right) =\phi_1 \left( p\right)
+\phi_2 \left( \mathbf{k}_{\Upsilon }\right)$. $\Box$
\vspace*{0.5\baselineskip}

The generation function $G(s,p,\mathbf{k}_{\Upsilon })$ has no singularities in
the region $\mathrm{Re\,}s,\mathrm{Re\,}p>0$. Thereby the left hand-side of
\eqref{geq} is equal to zero when $e^{s}-\Phi \left( p,\mathbf{k}_{\Upsilon
}\right) =0$. Resolving the latter equality with respect to the variable $p$ we
obtain a function $p=\varpi (s,\mathbf{k}_{\Upsilon })$ defined by the equation
\begin{equation}
    \Phi \left[ \varpi (s,\mathbf{k}_{\Upsilon }),\mathbf{k}_{\Upsilon }\right]
    = e^{s}
    \label{app20:eq.20}
\end{equation}
which specifies the locus in the space $\{s,p,\mathbf{k}_{\Upsilon }\}$ where
also the right hand-side of equation~\eqref{geq} has to be equal to zero. The
latter enables us to write immediately the boundary generation function in the
form
\begin{equation}
    g(s,\mathbf{k}_{\Upsilon }) =
    \frac{\exp \left[ -\varpi (s,\mathbf{k}_{\Upsilon})
    n_{0}\right] }{1-e^{-s}\phi \left[ \varpi (s,\mathbf{k} _{\Upsilon}),
    \mathbf{k}_{\Upsilon }\right]}\,.
    \label{app20:eq.21}
\end{equation}
Expressions \eqref{geq} and \eqref{app20:eq.21} actually solve the problem
giving us the following expression for the generation function
\begin{multline}
    G\left( s,p,\mathbf{k}_{\Upsilon }\right) =
    \frac{1}{1-e^{-s}}
\\
    \shoveleft{
    {}+
    \frac1{
    \left[ e^{s}-\Phi\left( p,\mathbf{k}_{\Upsilon }\right)\right]}
    \,\Bigg\{
    \frac{\left[ \Phi\left( p,\mathbf{k}_{\Upsilon }\right) -1\right] }
    {\left[ 1-e^{-s}\right]}
    }
\\
    {}+
    e^{
    -\left[ \varpi (s,\mathbf{k}_{\Upsilon })-p\right] n_{0}
    }
    \frac{
    \left[
    \phi \left( p,\mathbf{k}_{\Upsilon }\right) -\Phi
    \left( p,\mathbf{k}_{\Upsilon }\right) \right] }
    {
    \left[ 1-e^{-s}\phi
    \left( \varpi (s,\mathbf{k}_{\Upsilon }),\mathbf{k}_{\Upsilon }\right) \right] }
    \Bigg\}
    \label{gsolut}
\end{multline}
where the first summand is the image of the delta function
$\mathcal{P}_{t,\mathbf{n}}=\delta _{\mathbf{nn}_{0}}$ not contributing into
one of the quantities~\eqref{app2.v1}--\eqref{app2.v3}, the second term is due
to random walks over the internal nodes, and the last one is caused by the
boundary effects. Formula~\eqref{gsolut} specifies the desired generation
function in the general form.

\subsection{Limit of multiple-step random walks on small time scales}

In order to find the Laplace transforms~\eqref{app2.vl1}--\eqref{app2.vl3} it
suffices to expand the generation function $G\left( s,p,\mathbf{k}_{\Upsilon
}\right)$ into the Taylor series with respect to the arguments $p$ and
$\mathbf{k}_{\Upsilon }$ with cutting off the series at the second order terms.
However, in the case under consideration there are additional assumptions
simplifying essentially obtaining the desired results. First, only random walks
with many steps are of interest because the hop duration $\tau_a$ has been
chosen to be much less then the observation time interval $\tau $ of the
analyzed Markovian process, $\tau _{a}\ll \tau $. It means the inequality $s\ll
1$ to hold. Second, the time interval $\tau $ is regarded as any small value.
So only the components of moments~\eqref{app2.v1}--\eqref{app2.v3} that are
characterized by scaling $\tau ^{d}$ with the exponent $d$ not exceeding unity,
$d\leq 1$, are to be taken into account. With respect to the generation
function $G(s,p,\mathbf{k}_{\Upsilon })$ the latter assumption is converted to
the statement that all the components of itself and its derivatives calculated
at the point $\{\mathbf{k}_{\Upsilon }=\mathbf{0},\; p=0\}$ that scale with the
argument $s$ as $s^{-d}$ and have the exponent $d$ exceeding two, $d>2$, can be
ignored.

At the point  $\{\mathbf{k}_{\Upsilon }=\mathbf{0},~p=0\}$ according to their
definition \eqref{app20:eq.10}, \eqref{app20:eq.11} the function $\Phi (0,
\mathbf{0})=1$ and the function
\[
    \phi (0,\mathbf{0})=1-\frac{\sigma _{a}}{M},
\]
where the coefficient $\sigma _{a}$ is considered to be a small parameter,
which is justified in the limit $\tau_a\to0$ as will be seen below. Thereby in
the adopted assumptions expression~\eqref{gsolut} for the generation function
can be rewritten as
\begin{multline}
    G\left( s,p,\mathbf{k}_{\Upsilon }\right) =\frac{1}{s}+\frac{\left[ \Phi \left(
    p,\mathbf{k}_{\Upsilon }\right) -1\right] }{s^{2}}
\\
    {}+e^{-\varpi (s,
    \mathbf{0})n_{0}}\frac{\left[ \phi \left( p,\mathbf{k}_{\Upsilon }\right) -\Phi
    \left( p,\mathbf{k}_{\Upsilon }\right) \right] }{s\left[ s+1-\phi \left[ \varpi
    (s,\mathbf{0}),\mathbf{0}\right] \right] }.
    \label{gefun1}
\end{multline}
The expansion of the functions $\Phi \left( p,\mathbf{k}_{\Upsilon }\right) $,
$\phi \left( p,\mathbf{k}_{\Upsilon }\right) $  with respect to $p$ and
$\mathbf{k}_{\Upsilon }$ at the required order is
\begin{multline}
    \Phi (p,\mathbf{k}_{\Upsilon })  = 1  - \frac{\epsilon_M p}{M}
    +\frac{p^2}{2M}
\\
    {}+\frac{1}{M}\sum_{\alpha =1}^{M-1}
    \left(\mathrm{i}\epsilon _{\alpha }k_{\alpha }
    -\frac{1}{2}k_{\alpha }^{2}\right)
    \label{final:1}
\end{multline}
and
\begin{multline}
    \phi(p,\mathbf{k}_{\Upsilon }) = 1 -\frac{\sigma _{a}}{M}-\frac{p}{M}+
    \frac{p^{2}}{2M}
\\
    {}+\frac{\mathrm{i}g}{M}\sum_{\alpha =1}^{M-1}
    \epsilon _{\alpha }^{\Upsilon }k_{\alpha }
    -\frac{g}{2M}\sum^{M-1}_{\alpha,\beta =1}k_\alpha k_\beta\left(
    \delta_{\alpha\beta} +
    \frac{g-1}{M-1}\,\epsilon_{\alpha }^{\Upsilon }\epsilon _{\beta }^{\Upsilon }
    \right).
    \label{final:2}
\end{multline}
In deriving expression~\eqref{final:2} formulae~\eqref{bcor1}, \eqref{bcor2}
have been used. The substitution of the generation function written in
form~\eqref{gefun1} with approximations~\eqref{final:1}, \eqref{final:2} into
relations~\eqref{app2.vl1}--\eqref{app2.vl3} yields
\begin{align}
    \mathfrak{R}_{a}(s,n_{0}) &=\frac{\sigma
    _{a}}{M}\mathcal{K}_a\left(s,n_{0}\right)\,,
    \label{final:3}
\\
    \mathfrak{U}_{a}^{M}(s,n_{0}) &=
    \frac{1}{M}\mathcal{K}_a\left( s,n_{0}\right)+\frac{\epsilon _{M}}{Ms^{2}}\,,
    \label{final:4}
\\
    \mathfrak{U}_{a}^{\alpha }(s,n_{0}) &=
    \frac{(g-1)\epsilon _{\alpha }^{\Upsilon }}{M}\mathcal{K}_a
    \left( s,n_{0}\right)
    +\frac{\epsilon _{\alpha }}{Ms^{2}}\,,
    \label{final:5}
\\  \nonumber
    \mathfrak{L}_{a}^{\alpha \beta }(s,n_{0}) &=
    \frac{(g-1)}{2M}\left[ \delta _{\alpha \beta }+\frac{g
    \epsilon_{\alpha}^{\Upsilon }\epsilon _{\beta }^{\Upsilon }}{M-1}\right]
    \mathcal{K}_a\left( s,n_{0}\right)
    \\
    &\qquad{}+
    \frac{\delta _{\alpha \beta }}{2Ms^{2}}\,,
    \label{final:6}
\\
    \mathfrak{L}_{a}^{MM}(s,n_{0}) &=\frac{1}{2Ms^{2}}\,,
    \label{final:7}
\end{align}
the mean $\mathfrak{L}_a^{\alpha M}(s,n_0)$ is equal to zero. Here the function
$\mathcal{K}_a\left( s,n_{0}\right) $ is defined by the expression
\begin{equation}
    \mathcal{K}_a\left( s,n_{0}\right) =\frac{\exp\left[-\varpi (s,\mathbf{0})n_{0}\right]}
    {s\left[ s+1-\phi \left( \varpi (s,\mathbf{0}),\mathbf{0}\right)\right] }
    \label{final:10}
\end{equation}
and we have ignored some insignificant terms where appropriate.

Previously in the given appendix we measured time $t$ in units of the hop
duration $\tau_a$ and spatial coordinates $\{\zeta^i\}$ in units of the lattice
spacings $\{a_i\}$ within the frame $\mathfrak{b}$. Now let us return to the
initial units and deal with the corresponding spatial correlations. To do this,
first, functions~\eqref{final:4}--\eqref{final:7} should be multiplied by the
spacings $a_M$ and $a_\alpha$, or their products $a_\alpha a_\beta$ and
$a_M^2$, respectively. Second, the dimensionless Laplace argument $s$ has to be
replaced by the product $s\tau_a$, because previously when applying to the
discrete Laplace transformation the replacement
\[
         st\rightarrow s\tau_a\cdot\frac{t}{\tau_a}
\]
has be used obliquely. Third, for further converting the discrete Laplace
transformation into continuous one within the replacement
\[
       \tau_a\sum^{\infty}_{t/\tau_a = 0} \rightarrow \int^\infty_0 dt (\dots)
\]
all the functions~\eqref{final:3}--\eqref{final:7} must be multiplied by the
time scale $\tau_a$.

Leaping ahead we note that the absorption coefficient $\sigma_a$ has to scale
with $\tau_a$ as $\sigma_a\propto \sqrt{\tau_a}$. As before noted the
coefficients $\{\epsilon_i\}$ also behave in this way. Therefore the
observation time interval $\tau$ can be chosen to be so small that the solution
of equation~\eqref{app20:eq.20} become
\begin{equation}\label{final:11}
 \varpi(s\tau_a,\mathbf{0}) = \sqrt{2Ms\tau_a}
\end{equation}
and function~\eqref{final:10} matches a continuous Laplace transform
\begin{gather}
    \label{final:12}
    \tau_a \mathcal{K}_a(s\tau_a,n_{0}) = \sqrt{\frac{M}{2\tau_a}} \mathcal{K}(s,\zeta_0)
\\
\intertext{given by the expression}
    \label{final:13}
    \mathcal{K}(s,\zeta^M_0) =
    s^{-3/2} \exp\bigg(-\zeta_0\sqrt{\frac{s}{\mathcal{D}_M}}\;\bigg)\,,
\end{gather}
with $\zeta^M_0 = a_Mn_0$ being the distance from the node of the walker
initial position to the medium boundary $\Upsilon$ along the vector
$\boldsymbol{b}_M$.

Indeed, first, if we ignore the second term on the right-hand side of
expansion~\eqref{final:1} the solution of equation~\eqref{app20:eq.20} for
$s\tau_a\ll1$ and $\mathbf{k}_\Upsilon = \mathbf{0}$ is of
form~\eqref{final:11}. It is justified when $\varpi\gg \epsilon_M$, which is
equivalent to the condition $s\gg v^2_M/\mathcal{D}_M$ or $\tau\ll
\mathcal{D}_M/v^2_M$. Second, according to expansion~\eqref{final:2} the
denominator in expression~\eqref{final:10} at the leading order is
\[
    \left[ s\tau_a+1-\phi \left( \varpi (s\tau_a,\mathbf{0}),\mathbf{0}\right)\right]
    = \frac{\varpi(s\tau_a,\mathbf{0})}{M}
\]
provided $\varpi(s\tau_a,\mathbf{0})\gg \sigma_a$. Because $\sigma_a\sim
\sqrt{\varepsilon\tau_a}$, where $\varepsilon$ is some constant, the latter
inequality is reduced to the following $s\gg\varepsilon$ and
$\tau\ll\varepsilon$. Since the time interval is an arbitrary small value the
two inequalities can be adopted beforehand. Whence formulae~\eqref{final:11}
and \eqref{final:12} follows immediately for the spacing $a_M$ given by
expression~\eqref{app2.32b}.

\subsection{Continuum limit and a $\delta$-boundary model}

To get the final results we analyze the obtained expression in the limit
$\tau_a\to0$. The probability distribution $\mathcal{P}_{t,\mathbf{m}}$ of the
lattice random walks can be treated as the discrete implementation of the Green
function $G(\mathbf{r},\mathbf{r}_0,t)$ giving the probability density to find
a walker at the point $\mathbf{r}$ at time $t$ provided it was initially at the
point $\mathbf{r}_0$. Using the Green function $G(\mathbf{r},\mathbf{r}_0,t)$
the means under consideration are written as the following moments
\begin{align}
    \mathfrak{R}(t,\zeta_0) &=1-
    \int\limits_{\mathbb{R}^{M+}}d\mathbf{r}G(\mathbf{r},\mathbf{r}_0,t) \,,
    \label{cont:1}
    \\
    \mathfrak{U}_b^{i}(t,\zeta_0) &=\int\limits_{\mathbb{R}^{M+}}d\mathbf{r}
    (\zeta^{i}-\zeta_{0}^{i}) G(\mathbf{r},\mathbf{r}_0,t)\,,
    \label{cont:2}
                    \\
    \mathfrak{L}_b^{ij}(t,\zeta_0) &=\frac12
    \int\limits_{\mathbb{R}^{M+}}d\mathbf{r}
    (\zeta^{i}-\zeta_{0}^{i}) (\zeta^{j}-\zeta_{0}^{j})
    G(\mathbf{r},\mathbf{r}_0,t)\,,
    \label{cont:3}
\end{align}
and their Laplace transforms can be obtained from the
quantities~\eqref{final:3}--\eqref{final:7} in the manner described in the
previous subsection. As the result we have
\begin{align}
    \mathfrak{R}(s,\zeta_{0}) &=D_{MM}^{-1/2}\sigma\,\mathcal{K}(s,\zeta_{0})\,,
    \label{cont:4}
\\
    \mathfrak{U}_{b}^{M}(s,\zeta_{0}) &=
    D_{MM}^{-1/2}\omega\,\mathcal{K}( s,\zeta_{0})+\frac{v^{M}}{s^{2}}\,,
    \label{cont:5}
\\
    \mathfrak{U}_{b}^{\alpha }(s,\zeta_{0}) &=
    D_{MM}^{-1/2}l_\Upsilon v_\Upsilon^{\alpha}\,
    \mathcal{K} ( s,\zeta_{0})
    +\frac{v^{\alpha }}{s^{2}}\,,
    \label{cont:6}
\\
    \mathfrak{L}_{b}^{\alpha \beta }(s,\zeta_{0}) &=\Big[
    D_{MM}^{-1/2}l_\Upsilon \mathcal{D}_\alpha \,\mathcal{K}( s,\zeta_{0})
    +
    \frac{\mathcal{D}_\alpha}{s^{2}}\Big]\delta _{\alpha \beta }\,,
    \label{cont:7}
\\
    \mathfrak{L}_{b}^{MM}(s,\zeta_{0}) &=\frac{\mathcal{D}_M}{s^{2}}
    \label{cont:8}
\end{align}
the component $\mathfrak{L}_b^{\alpha M}(s,\zeta_0)$ is equal to zero. Here the
following characteristics of the medium boundary treated as an infinitely thin
layer $\Upsilon$
\begin{align}\label{cont:9}
    \sigma &:= \sigma_a \sqrt{\frac{D_{MM}}{2M\tau_a}}\,,
    & l_\Upsilon& := g\sqrt{\frac{MD_{MM}\tau_a}{2}}
\end{align}
have been introduced and expression~\eqref{app2.Prep2.D4} have been used. It
should be noted that according to \eqref{cont:9} the number $g$ of elementary
hops forming the long distant jumps of wallers in the boundary layer
$\Upsilon_0$ has to grow with $\tau_a$ as $\tau_a^{-1/2}$ in order to retain
the effect of boundary fast transport in the limit $\tau_a\to0$. As a result,
the second term in the square brackets of expression~\eqref{cont:5} scales as
$\sqrt{\tau_a}$ because, in turn, the coefficients
$\{\epsilon^\Upsilon_\alpha\}$ vary with $\tau_a$ as $\sqrt{\tau_a}$. Therefor
it vanishes in the limit $\tau_a\to0$ and the symmetry of the second moments
caused by the boundary fast diffusion is restored.

The equality (see, e.g., Ref.~\cite{IntegTab})
\begin{equation*}
    \int\limits_0^\infty \frac{dt}{\sqrt{\pi t}}\exp\left(
    -\frac{\zeta_0^2}{4\mathcal{D}_M t}-st   \right) =
    \frac{1}{\sqrt{s}}\exp\left(-\zeta_0\sqrt{\frac{s}{\mathcal{D}_M}}\;\right)
\end{equation*}
and the Laplace transform of integrals enable us represent the inverse Laplace
transform $\mathcal{K}(t,\zeta_0)$ of function~\eqref{final:13} in the integral
form
\begin{equation}\label{cont:10}
    \mathcal{K}(t,\zeta_0) = \sqrt{\frac{t}{\pi}} \int\limits_0^1
    \frac{dz}{\sqrt{z}}
    \exp\left(-\frac{\zeta_0^2}{4\mathcal{D}_M t}\,\frac1z\right)\,.
\end{equation}
Expression~\eqref{cont:10} together with
formulae~\eqref{cont:4}--\eqref{cont:8} proves Proposition~\ref{app2.Prep3}.

\begin{acknowledgements}
The authors are grateful to V.G. Morozov for the discussion of the obtained
results. This work has been supported in part by DFG project MA 1508/8-1 and
RFBR grants 06-01-04005, 05-01-00723, 05-07-90248, and 04-02-81059.
\end{acknowledgements}

\end{document}